\documentclass[journal]{IEEEtran}
\usepackage{CJKutf8}
\usepackage{amsmath,amssymb,amsfonts}
\usepackage{algorithm}
\usepackage{algorithmic}
\usepackage{array}
\usepackage{textcomp}
\usepackage{booktabs}
\usepackage{stfloats}
\usepackage{url}
\usepackage{verbatim}
\usepackage{graphicx}
\usepackage{cite}
\usepackage{xcolor}
\usepackage{makecell}
\usepackage{multirow}
\usepackage[font=small]{caption} 
\usepackage{orcidlink} 
\hypersetup{hidelinks} 
\usepackage{hyperref}
\hypersetup{hypertex=true,
            colorlinks=true,
            linkcolor=blue,
            anchorcolor=blue,
            citecolor=blue}
\def\BibTeX{{\rm B\kern-.05em{\sc i\kern-.025em b}\kern-.08em
    T\kern-.1667em\lower.7ex\hbox{E}\kern-.125emX}}

\begin{document}
\begin{CJK}{UTF8}{gbsn}

\title{Data Augmentation for Seizure Prediction\\ with Generative Diffusion Model}

\author{Kai Shu$^{\orcidlink{0009-0007-8707-0085}}$, Le Wu$^{\orcidlink{0000-0002-8565-9626}}$,~\IEEEmembership{Member,~IEEE,} Yuchang Zhao$^{\orcidlink{0000-0002-1726-7962}}$, Aiping Liu$^{\orcidlink{0000-0001-8849-5228}}$,~\IEEEmembership{Member,~IEEE,} Ruobing Qian, and Xun Chen$^{\orcidlink{0000-0002-4922-8116}}$,~\IEEEmembership{Senior Member,~IEEE} 
\thanks{This work was partially supported by the National Natural Science Foundation of China under Grants 32271431 and W2432042, the Research Project of Health Commission of Anhui Province under Grant AHWJ2022b004, and the Joint Fund for Medical Artificial Intelligence under Grant MAI2023C004. \emph{(Corresponding author: Xun Chen.)}}
\thanks{Kai Shu, Le Wu, Yuchang Zhao, Aiping Liu and Xun Chen are with the School of Information Science and Technology, University of Science and Technology of China, Hefei 230027, China (e-mail: xunchen@ustc.edu.cn).}
\thanks{Ruobing Qian is with the Epilepsy Center, Department of Neurosurgery, The First Affiliated Hospital of USTC, Division of Life Sciences and Medicine, University of Science and Technology of China, Hefei 230001, China.}} 

\markboth{IEEE TRANSACTIONS ON COGNITIVE AND DEVELOPMENTAL SYSTEMS}%
{Shell \MakeLowercase{\textit{et al.}}: Data Augmentation for Seizure Prediction with Generative Diffusion Model}


\maketitle

\begin{abstract}
Data augmentation (DA) can significantly strengthen the electroencephalogram (EEG)-based seizure prediction methods. However, existing DA approaches are just the linear transformations of original data and cannot explore the feature space to increase diversity effectively. Therefore, we propose a novel diffusion-based DA method called DiffEEG. DiffEEG can fully explore data distribution and generate samples with high diversity, offering extra information to classifiers. It involves two processes: the diffusion process and the denoised process. In the diffusion process, the model incrementally adds noise with different scales to EEG input and converts it into random noise. In this way, the representation of data can be learned. In the denoised process, the model utilizes learned knowledge to sample synthetic data from random noise input by gradually removing noise. The randomness of input noise and the precise representation enable the synthetic samples to possess diversity while ensuring the consistency of feature space. We compared DiffEEG with original, down-sampling, sliding windows and recombination methods, and integrated them into five representative classifiers. The experiments demonstrate the effectiveness and generality of our method. With the contribution of DiffEEG, the Multi-scale CNN achieves state-of-the-art performance, with an average sensitivity, FPR, AUC of 95.4\%, 0.051/h, 0.932 on the CHB-MIT database and 93.6\%, 0.121/h, 0.822 on the Kaggle database.
\end{abstract}


\section{Introduction}\label{Intr}
\IEEEPARstart{E}{pilepsy} is caused by abnormal discharge of brain neurons and is one of the world's most common neurological diseases \cite{fisher2014ilae}. It is reported by the World Health Organization (WHO) that about 50 million people are suffered from epilepsy. With the development of medical science, 70\% of the patients can be seizure-free after proper treatment. However, the remaining 30\% of them still cannot be controlled by drugs \cite{world2019epilepsy}. Therefore, seizure prediction has great value for them. A precise prediction can let patients get timely protection, improving the quality of their lives by reducing their mental stress.

Electroencephalogram (EEG) is an efficient method for capturing and monitoring electrical activity of the brain \cite{zeng2020hierarchy}. EEG recordings can be divided into three kinds depending on the position where the signals are collected. There is non-invasive scalp electroencephalogram (sEEG) with electrodes attached to the subject's scalp, semi-invasive electrocorticogram (ECoG) with electrodes placed between the cerebral cortex and the dura mater, and invasive electroencephalogram (iEEG) with implanted electrodes \cite{maimaiti2022overview}. EEG is widely used to diagnose epilepsy in clinical \cite{xiao2021automatic}. Over the last few years, studies have demonstrated that EEG can be used to predict upcoming seizures \cite{cao2019epileptic}\cite{kuhlmann2018seizure}. In practice, physicians have divided the epilepsy EEG signals into four states: preictal (the period before seizures), ictal (the duration of seizures), postictal (the period following seizures) and interictal state (the interval between seizures) \cite{daoud2019efficient}. According to the definitions above, seizure prediction is transferred to a binary classification task which distinguishes preictal states from interictal ones. In this way, a variety of methods have been proposed to achieve the prediction. 

In the traditional machine learning methods, many kinds of manually extracted features have been used to predict seizures, including spatial domain, time domain, frequency domain, and time-frequency domain features. After feature extraction, a classifier is used to do the classification by these features \cite{chen2022toward}. For instance, Chisci et al. utilized the autoregressive coefficient as features, and a support vector machine (SVM) as the classifier \cite{chisci2010real}. However, extracting features manually requires extensive expertise, and the efficacy of seizure prediction is heavily reliant on the quality of these extracted features. Consequently, deep learning (DL) methods have gained growing attention within the last few years, where representative features can be extracted automatically instead of manually \cite{ozcan2019seizure}\cite{cao2021epileptic}. Researchers have designed various DL architectures to achieve high-performance seizure prediction. Li et al. applied the multi-layer perceptrons (MLP) blocks to extract spatial and temporal features \cite{li2023spatio}. Gao et al. employed a multi-scale convolutional neural network (CNN) by using the dilated convolutions to capture information at different levels \cite{gao2022pediatric}. In another work \cite{gao2022general}, Gao et al. proposed a Transformer-based network. The model utilized the attention mechanism to establish associations between information at different locations and strengthen the role of key information. The multiple features extracted by these networks had great effects on seizure prediction and these DL methods all obtained satisfactory performance.

It is worthy noting that DL algorithms typically need a substantial amount of labeled data to achieve good performance. However, the low-frequency property of seizure onsets causes serious insufficiency of the preictal data for a specific patient \cite{cook2013prediction}\cite{ihle2012epilepsiae}. Thus, the composition of dataset is seriously imbalanced, leading to the overfitting during model training \cite{branco2016survey}.            

To solve the issue of imbalanced data, various methods have been employed. Khan et al. selected interictal samples randomly to down-sample the interictal data \cite{khan2017focal}. However, down-sampling discards too many interictal samples, which could lead to the loss of useful information. Besides, the small quantity of data after down-sampling makes it hard for models to reach the optimal result. Therefore, some researchers proposed data augmentation (DA) methods. DA is the process that generates new samples to augment a small or imbalanced dataset. Troung et al. generated more preictal segments by using an overlapping technique which slid a 30-s window along the EEG signal \cite{truong2018convolutional}. Zhang et al. came up with an idea of recombination. For each seizure, they split every training EEG sample into three segments, and generated new samples as a recombination of the randomly selected segments \cite{zhang2019epilepsy}. Although those traditional DA methods can increase the number of preictal samples, they are just simple transformations of the existing data, with distribution constrained by original data due to the presence of numerous repetitive parts \cite{antoniou2017data}. Because of the nonstationary signals in the brain, the epileptic representation of EEG varies over time \cite{qi2021learning} and an upcoming seizure may have representation biased from the previous ones. The simple transformations of existing samples can hardly expand the data distribution to cover the variable representation.

With the advancement of generative model, an increasing number of researchers have utilized these models to address the imbalanced data \cite{trabucco2023effective}. Rasheed et al. used a generative adversarial network (GAN) to produce synthetic EEG feature maps and evaluated the performance of generated data by a validation method \cite{rasheed2021generative}. However, the training process of GAN is difficult. Once the design is improper, the gradient may disappear or explode, resulting in unstable generation quality. In addition, the model may fall into local minimum during training and can only generate samples similar to a limited number of initial data. This is known as the mode collapse problem and will lead to the lack of generation diversity \cite{salimans2016improved}.

\begin{figure}[htbp]
	\centerline{\includegraphics[width=0.5\textwidth]{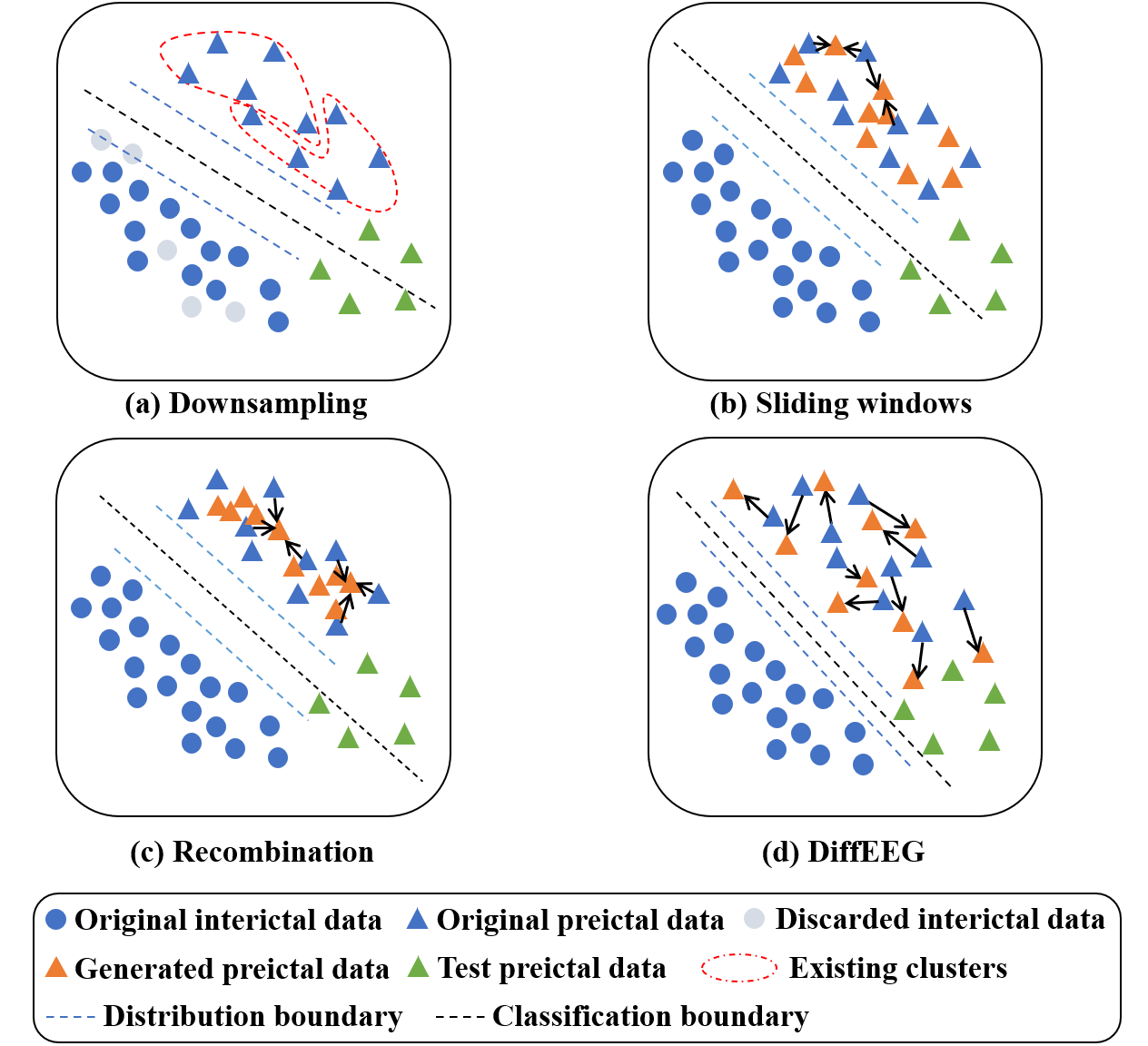}}
	\caption{The distribution illustration of down-sampling, sliding windows, recombination and DiffEEG. Each cluster represents the preictal data of a seizure. The clusters of different preictal data exhibit both similar and distinct distribution. (a) Down-sampling of interictal samples may cause the loss of useful information. (b) When using the sliding windows, the distribution of generated samples is between the front and rear samples. (c) The recombination is implemented within each seizure, because samples recombined by segments from different seizures have poor authenticity. The distribution of recombined samples is at the center of the three component samples and thus limited by them. (d) Instead, DiffEEG could fully explore the feature space and expand the distribution to outward area. The generated samples connect different clusters into a whole.}
	\label{fig1}
\end{figure}

Recently, diffusion models have been widely used in various fields such as computer vision (CV) and natural language processing (NLP) \cite{giannone2022few}\cite{kong2020diffwave}. Diffusion models are a class of promising generative models, which regard data as a step-by-step diffusion of a series of random noises. They use a Markov chain to gradually add Gaussian noise to data, getting the posterior distribution probability and learning the reverse denoised process. Thus they could synthesise new samples from random noise with the learned distribution of original data \cite{ho2020denoising}. One important use of the diffusion model is DA due to its strong generation ability \cite{kebaili2023deep}\cite{pinaya2022brain}\cite{chung2022mr}. For example, Chambon et al. utilized a pre-trained latent diffusion model to generate high-fidelity, diverse synthetic chest x-rays (CXR) and measured a 5\% improvement by jointly training a classifier on synthetic and real images \cite{chambon2022roentgen}. In this paper, we proposed a diffusion-based model named DiffEEG. In specific, DiffEEG consists of two opposite processes, namely the forward/diffusion process and the reverse/denoised process. In the diffusion process, the model adds noise with different scales to the EEG sample step by step and converts it into random noise. This procedure is conditioned on the short-time Fourier transform (STFT) spectrogram of the input EEG to provide the guiding time-frequency features. By minimizing the loss between the output of the network and the added noise, the feature distribution of preictal data could be fully explored. In the denoised process, based on the learned conditional probability distribution, the synthetic data could be sampled from random noise input by gradually removing the noise with different scales. With the randomness of initial noise and the precise representation, the generated samples could have new information and diversity while ensuring the consistency of feature space, narrowing the distance between data clusters composed by initial samples from each seizure \cite{dhariwal2021diffusion}. The distribution illustration of down-sampling, sliding windows, recombination and DiffEEG is presented in Fig. \ref{fig1}. Different from the existing DA methods, DiffEEG is capable of utilizing the learned distribution information to generate samples of diversity and provide additional information for model training.

We train DiffEEG with the patient's preictal EEG signals so that our model can generate synthetic preictal data until we have an equal number of samples per class. The generation quality is improved with the reduction of the training loss. The samples are then given to a classifier to conduct seizure prediction task. We used five representative network frameworks (SVM, Spatio-temporal MLP, EEGNet, Multi-scale CNN, Transformer) as classifiers on the CHB-MIT and Kaggle dataset to verify the universality and generality of our model. In addition, we compared our model with that using just the original data and those using three existing methods for solving the data imbalance, including down-sampling, sliding windows and recombination, to evaluate the effectiveness and superiority of DiffEEG. 

The main contributions of our work are as follows:
\begin{itemize}
	\item We propose a method of using diffusion model to solve the imbalanced data problem, which is the first work to introduce diffusion model to the seizure prediction field. 
	\item We verify that our diffusion model is superior to the existing DA methods and can be utilized in almost any other classification network to substantially improve the seizure prediction performance.  
	\item With the improvement of DiffEEG, the best prediction performance in our experiments outperforms that of most state-of-the-art (SOTA) methods. 
\end{itemize}

The rest of the paper is structured as follows. Section \ref{Meth} details our proposed methods, including the theory of diffusion model, the structure of DiffEEG and five classifiers. Section \ref{Expe} shows the datasets, our experimental results and comparison with SOTA works. Section \ref{Disc} discusses the results of our methods. Finally, we conclude our work in Section \ref{Conc}.

\section{Methodology}\label{Meth}
\subsection{Diffusion Model}
\begin{figure*}[htbp]
	\centerline{\includegraphics[width=0.85\textwidth]{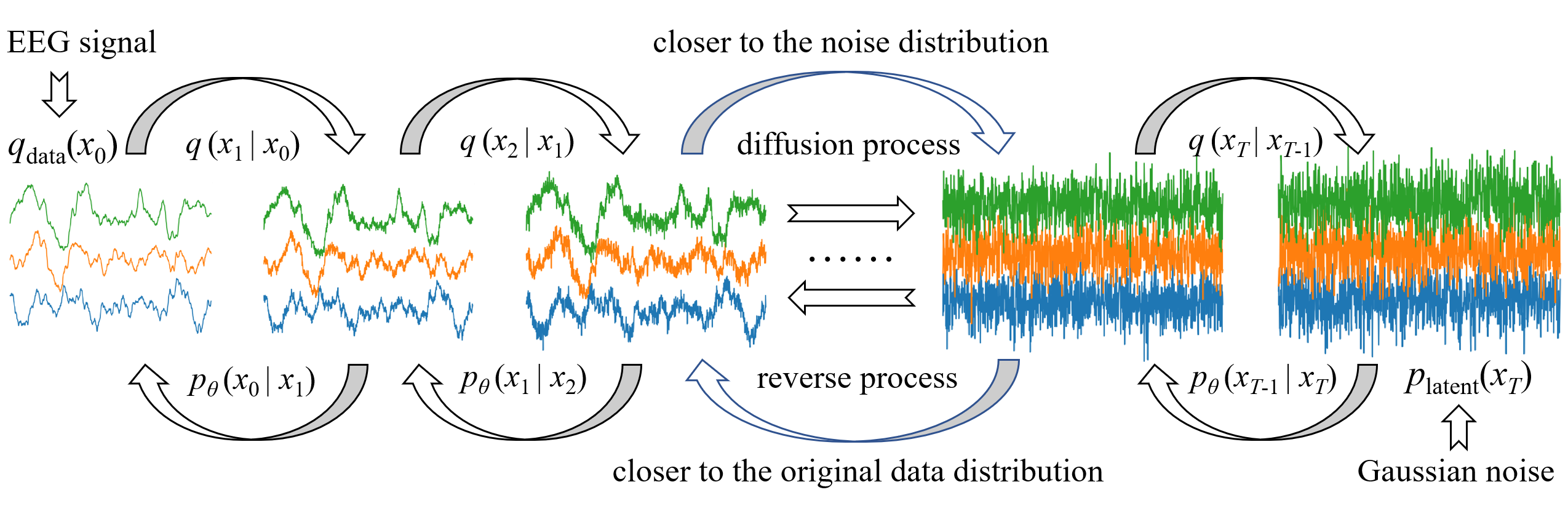}}
	\caption{The diffusion and reverse process in diffusion model.}
	\label{fig2}
\end{figure*}

Since its inception, the diffusion model has shown its remarkable advantages in various fields, offering a unified loss function with high simplicity and a stable training process with adequate theoretical support \cite{cao2022survey}. The diffusion model aims to transform the prior data into random noise and revise the transformations step by step to rebuild a brand new sample with the same distribution as the prior. We define $x_t\in R^{H \times L}$ for $t$ = 0, 1, 2, ... , $T$ as the data at diffusion step $t$, where $H$ is the EEG channel and $L$ is the length of EEG segment. The process that transforms the starting state $x_0$ into random noise $x_T$ through a Markov chain $q(x_t|x_{t-1})$ is the forward/diffusion process:
\begin{equation}
	q(x_1,...\ , x_T|x_0) = \prod_{i=1}^{T} q(x_t|x_{t-1}).
\end{equation}
The process that converts random noise $x_T$ to data $x_0$ with the starting distribution through a Markov chain $p_{\theta}(x_{t-1}|x_t)$ is called the reverse/denoised process, which is parameterized by $\theta$:
\begin{equation}
	p_{\theta}(x_0,...\ ,x_{T-1}|x_T) = \prod_{i=1}^{T} p_{\theta}(x_{t-1}|x_t).
\end{equation}
The diffusion and reverse process are showed in Fig. \ref{fig2}. Given $x_T \sim N(0, I)$, we can sample $x_{t-1} \sim p_{\theta}(x_{t-1}|x_t)$ step by step, until we get the final synthetic data $x_0$ with the same distribution as real data. 

Based on the variance schedule $\{\beta_t\}_{t=1}^T$ of the noise added during the diffusion process, some constants about noise can be further defined:
\begin{equation}
  \alpha_t = 1-\beta_t,\  \overline{\alpha}_t=\prod_{s=1}^{t}\alpha_s,\ 
  \widetilde{\beta}_t = 
  \begin{cases}
    \beta_1, & t = 1 \\
    \frac{1-\overline{\alpha}_{t-1}}{1-\overline{\alpha}_t}\beta_t, & t > 1 \\ 
  \end{cases}.
\end{equation}

We then define the mean $\mu_{\theta}$ and standard deviation $\sigma_{\theta}$ of the conditional probability distribution from $x_t$ to $x_{t-1}$:
\begin{gather}
     \mu_{\theta}(x_t,t) = \frac{1}{\sqrt{\alpha_t}}(x_t-\frac{\beta_t}{\sqrt{1-\overline{\alpha}_t}}\varepsilon_{\theta}(x_t,t)), \notag \\
     \sigma_{\theta}(x_t,t) = \sqrt{\widetilde{\beta}_t}.
\end{gather}

$\varepsilon_{\theta}: R^{H \times L} \rightarrow R^{H \times L}$ is the diffusion network, with inputs $x_0$, Gaussian noise $\varepsilon \sim N (0, I)$ and diffusion step $t$. 

The training of diffusion model is to maximize its variational lower bound (ELBO):
\begin{equation}
	ELBO = E_{q(x_0,...,x_T)}log\frac{p_{\theta}(x_0,...\ ,x_{T-1}|x_T)\cdot p(x_T)}{q(x_1,...\ ,x_T|x_0)}.
\end{equation}

It is proved that high generation quality can be achieved by minimizing the unweighted variant of ELBO \cite{ho2020denoising}, which is used as the training objective of our model: 
\begin{equation}\label{equ:6}
	\min\limits_{\theta}L(\theta) = E_{x_0,\varepsilon,t}\|\varepsilon-\varepsilon_{\theta}(\sqrt{\overline{\alpha}_t}x_0+\sqrt{1-\overline{\alpha}_t}\varepsilon,\ t)\|_2^2 .
\end{equation}

The training and sampling procedures are in Algorithm \ref{alg:1} and \ref{alg:2}, respectively.

\begin{algorithm}[H]
\caption{Training}
\begin{algorithmic}
\label{alg:1}
\REPEAT {
\STATE	Sample $x_0 \sim q_{data}$, $\varepsilon \sim N (0, I)$, and\\
    $t$ $\sim$ Uniform(\{1, 2, ... , $T$\})\\
	Take gradient descent step on\\ $\nabla_{\theta}\|\varepsilon-\varepsilon_{\theta}(\sqrt{\overline{\alpha}_t}x_0+\sqrt{1-\overline{\alpha}_t}\varepsilon,\ t)\|_2^2$
}
\UNTIL {converged}
\end{algorithmic}
\end{algorithm}

\begin{algorithm}[H]
\caption{Sampling}
\begin{algorithmic}
\label{alg:2}
\STATE	Sample $x_T \sim N(0, I)$
\FOR {$t$ = $T$, $T$-1, ... , 1}{
\STATE	Compute $\mu_{\theta}$($x_t$,t) and $\sigma_{\theta}$($x_t$,t)\\
	Sample $x_{t-1} \sim p_{\theta}(x_{t-1}|x_t)$ = \\
	N($x_{t-1};\mu_{\theta}(x_t,t),\sigma_{\theta}(x_t,t)^2I$)
}
\ENDFOR
\end{algorithmic}
\end{algorithm}

According to Algorithm \ref{alg:2}, given $x_T \sim N(0, I)$, $x_{T-1}$ can be generated by sampling the standard Gaussian noise $z$:
\begin{equation}
        x_{T-1} = \frac{1}{\sqrt{\alpha_T}}(x_T-\frac{\beta_T}{\sqrt{1-\overline{\alpha}_T}}\varepsilon_{\theta}(x_T,T)) + \sigma_{\theta}(x_T,T) \cdot z.
\end{equation}
$\frac{\beta_T}{\sqrt{1-\overline{\alpha}_T}}\varepsilon_{\theta}(x_T,T)$ is the Gaussian noise removed in the first step, which is estimated by the trained diffusion model. It's worth noting that only the combination of mutually independent Gaussian distributions can be Gaussian, but the removed noise is related to $x_T$. As a consequence, $x_{T-1}$ is not necessarily Gaussian. In this manner, the model continues to sample $x_{T-2}$, $x_{T-3}$, ... , $x_0$. The final generated sample $x_0$ may not necessarily be Gaussian either. Hence, the performance of the diffusion model is not affected by whether the data distribution is Gaussian. The model can adapt to generation tasks under any data distribution and synthesis samples conforming to the corresponding distribution, which is one of the reasons why it is widely used in fields such as CV, NLP and others. 

In order to shorten the training and sampling time, acceleration strategies are adopted in both the training and inference processes. During the training process, a random-step training strategy is utilized, which has been widely used since its initial proposal by Ho et al. in Denoising Diffusion Probabilistic Models (DDPM) \cite{ho2020denoising}. Based on the properties of the Gaussian distribution, $x_t$ can be obtained from $x_0$ by adding noise in one step:
\begin{equation}
        x_t =  \sqrt{\overline{\alpha}_t}x_0 + \sqrt{1-\overline{\alpha}_t}{\epsilon}.
\end{equation}

Compared with full-step training, random-step training substantially reduces the computational overhead while ensuring that the model has the opportunity to learn the data distribution at all diffusion steps.

The acceleration strategy used in the inference process is the skip-step sampling scheme proposed by Song et al. in Denoising Diffusion Implicit Models (DDIM) \cite{song2020denoising}. In order to implement skip-step sampling, it is necessary to construct a non-Markovian noise-adding process. Song et al. proved that the loss function of this process is
\begin{gather}
    L = KL(p(x_{t-1}|x_t)||p_{\theta}(x_{t-1}|x_t)) \propto  \notag \\
    \frac{(\sqrt{1-{\alpha}_{t-1}-{\sigma}_t^2}-\sqrt{\frac{{\alpha}_{t-1}(1-{\alpha}_t)}{{\alpha}_t}})^2}{2{\sigma}_t}||{\epsilon}_{\theta}(x_t,t)-{\epsilon}_t||^2,
\end{gather}
which remains consistent with equation (\ref{equ:6}), the loss function of the Markovian form in DDPM. This allows the continue utilization of the objective function from DDPM to train the network, while the inference process, not bound by the Markovian property, can employ the skip-step method to accelerate the sampling speed.

\subsection{DiffEEG Architecture}
The raw EEG signals are characterized by serious non-stationarity. Their spectral composition keeps varying over time, making it difficult for models to learn the representation in the time or frequency domain alone \cite{khan2018new}. The short-time Fourier transform (STFT) is extensively applied in the analysis of non-stationary signals. It can reflect the temporal variations of the spectral component and capture the epileptic representation of EEG signals. As a matter of fact, many studies have adopted STFT instead of raw EEG signals as inputs to the classification networks for seizure prediction \cite{truong2018convolutional} \cite{yang2021effective}. Therefore, we utilized the STFT spectrum in our DiffEEG model as the conditional guidance. 

\begin{figure*}[htbp]
	\centerline{\includegraphics[width=0.85\textwidth]{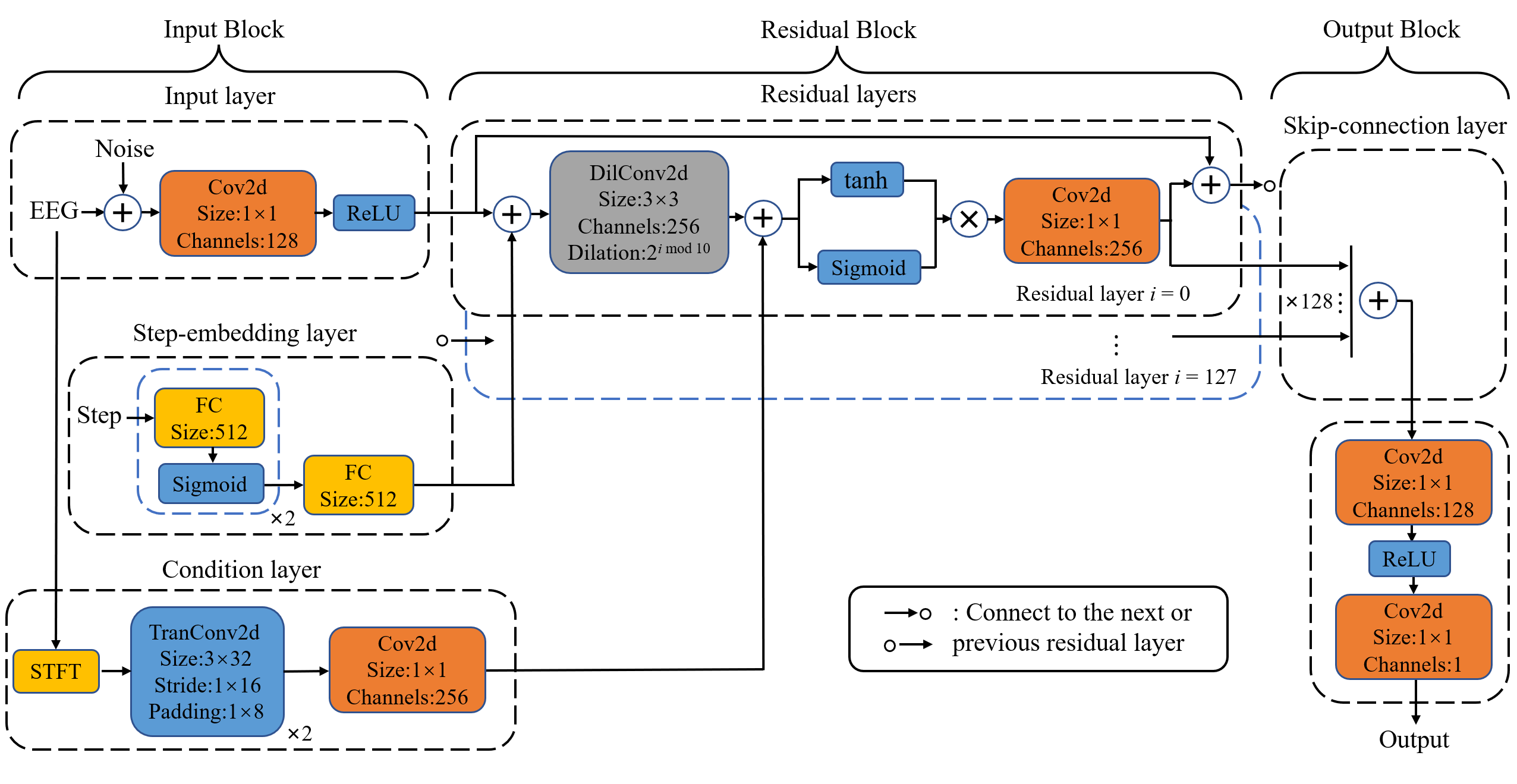}}
	\caption{DiffEEG Architecture.}
	\label{fig3}
\end{figure*}

Based on the characteristic of multi-channel EEG data, we have proposed the diffusion-based network DiffEEG with conditional guidance. The network is composed of the input block, the residual block and the output block. Fig. \ref{fig3} shows the whole architecture.

The input block consists of three layers. The input layer adds noise to the input EEG data to diffuse it forward by one step, and uses a 1×1 convolution (Conv) with $C$ kernels to map the data into $C$ residual channels. The 1×1 Conv is a way to increase the non-linearity of the network without affecting the receptive fields \cite{simonyan2014very}. It is utilized to raise the channel dimension while preserving the original shape of the data. The step-embedding layer is to encode the current diffusion-step and transform it into a $C$-dimensional embedding vector. As the added noise and the output of the model vary with different diffusion-step, it is necessary to inject the step information into the network. The sine and cosine functions mentioned in \cite{vaswani2017attention} are applied to convert the step $t$ into a learnable embedding $t_{eb}$, as showed in equation (\ref{equ:10}). These functions are selected due to the property that for any given offset $k$, $t_{eb}(x)$ can be expressed as a linear function of $t_{eb}(x+k)$, making it easy for the model to learn the relative positions between steps. Three fully connected (FC) layers are subsequently employed to project the embedding into the high-dimensional feature space and learn the step information. The condition layer utilizes the STFT spectrogram of the original signal as the conditional guidance. This condition is capable of providing time-frequency information for the network to understand the complex EEG rhythms, by learning the correlation between the fluctuations in EEG and the frequency content behind them. For CHB-MIT dataset, we set the STFT window length to 256 with no overlap while for Kaggle dataset the window length is 400 with 300 overlap. The STFT spectrogram is then up-sampled by two transposed 2-D convolutions to acquire the same shape as the EEG input in time domain. The transposed convolution can learn the up-sampling parameters that best suit the characteristics of the data. At last, there is a 1×1 Conv with 2$C$ kernels to merge the features across different frequencies. 
\begin{equation}\label{equ:10}
	\begin{split}
		t_{eb} = [\sin{(10^{\frac{0\times 4}{63}}t)},...\ ,\sin{(10^{\frac{63\times 4}{63}}t)},\\
		\cos{(10^{\frac{0\times 4}{63}}t)},...\ ,\cos{(10^{\frac{63\times 4}{63}}t)}]
	\end{split}
\end{equation}

The residual block contains $N$ residual layers with $C$ residual channels. In each residual layer, we employ a bidirectional dilated convolution (Bi-DilConv) \cite{oord2016wavenet} whose dilation is doubled between layers. The dilated convolution introduces intervals within the kernels and enlarges the sliding stride, thereby expanding the receptive field of each neuron without increasing the computational overhead. The exponential growth in dilation is to provide a multi-scale receptive field, learning both local and global information. Meanwhile, we take the modulus of dilation by 10 to prevent the exponential explosion. The diffusion-step embedding and STFT condition from the input block are added as the bias terms before and after the Bi-DilConv. Then a gated-tanh unit (GTU) is applied to learn the nonlinear features \cite{dauphin2017language}. The GTU is a combination of the tanh and sigmoid activation functions. The sigmoid activation acts as a gate for the tanh activation, regulating the weight of the tanh activation’s output. At the end of each residual layer is a 1×1 Conv with 2$C$ kernels to linearly combine the features from different channels. We take advantage of the residual structure in \cite{he2016deep}. The input of each residual layer is divided into two branches: one passes through the Bi-DilConv, GTU and 1×1 Conv as in a general network, while the other is directly connected to the output of the 1×1 Conv. The output of the 1×1 Conv is also divided into two branches: one is added with the input and connect to the next layer, while the other skips all the subsequent residual layers and is directly fed into the final output block. The purpose of this structure is to mitigate the vanishing gradient problem because the gradient of the input and output can propagate directly to the end of the network without traversing the intermediate layers.

The output block involves two layers. The skip-connection layer sums the outcomes from $N$ residual layers to integrate the features of different scales. The output layer utilizes two 1×1 convolutions to transform the data into the final output with the same dimensions as the input EEG data.

\subsection{Classifier Architecture}
We choose five representative SOTA network frameworks as classifiers. They are SVM, Spatio-temporal MLP \cite{li2023spatio}, EEGNet \cite{lawhern2018eegnet}, Multi-scale CNN \cite{gao2022pediatric} and Transformer \cite{gao2022general}.

\subsubsection{SVM}
SVM is a commonly used machine learning classifier with strong and stable classification capacity. It aims to find an optimal hyperplane that separates different classes of samples. Support vectors are the samples closest to the hyperplane, which play a crucial role in determining the position of hyperplane. By maximizing the distance between the support vectors and the hyperplane, SVM can provide powerful generalization and robustness. Before employing SVM, features need to be extracted manually from the EEG data. We adopted the features used in \cite{ozcan2019seizure}, including 4 statistical moments: mean, variance, skewness and kurtosis, 8 spectral band power: $\delta$ (0.5-4 Hz), $\theta$ (4-8 Hz), $\alpha$ (8-13 Hz), $\beta$ (13-30 Hz), $\gamma$-1 (30-50 Hz), $\gamma$-2 (50-75 Hz), $\gamma$-3 (75-100 Hz) and $\gamma$-4 (100-128 Hz), and 2 Hjorth parameters: mobility and complexity. 

\subsubsection{Spatio-temporal MLP}
Spatio-temporal MLP is composed by a preprocessing block and an MLP block. The preprocessing block contains three layers: a denoising layer to remove the muscle and ocular artifacts \cite{hu2015removal}, a weighted layer to strengthen effects of the key channels, and a reduction layer to reduce the computation cost. The MLP block involves an inter-channel layer that learns the spatial correlation between channels and an intra-channel layer that extracts information within channels. At last, there is an average pooling layer followed by the FC layer to obtain the probability of each seizure state.

\subsubsection{EEGNet}
\begin{figure*}[htbp]
	\centerline{\includegraphics[width=0.85\textwidth]{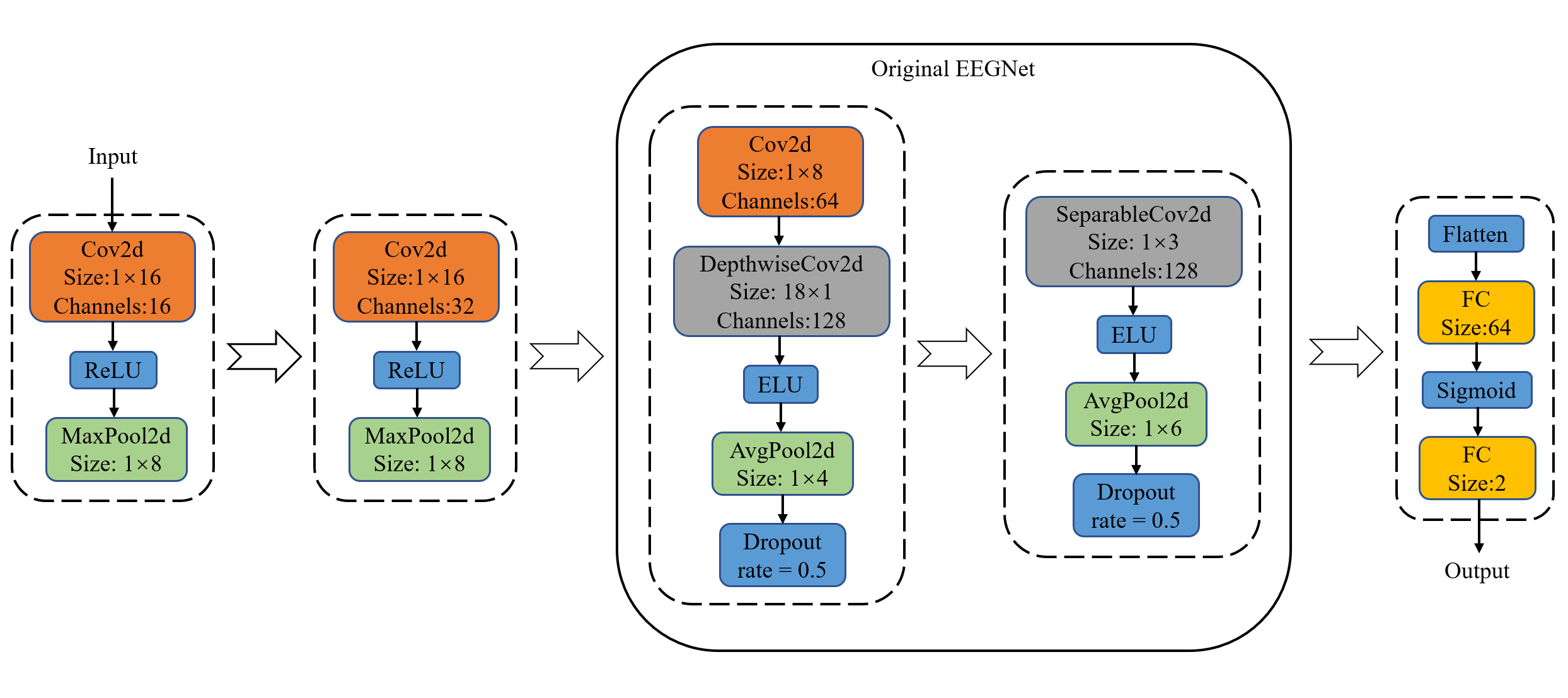}}
	\caption{The structure of the modified EEGNet.}
	\label{fig4}
\end{figure*}

EEGNet is a compact CNN that has been successfully applied to various EEG classification tasks \cite{lawhern2018eegnet}. There are two blocks in EEGNet. The first block starts with a convolution to capture the frequency information, followed by a depthwise convolution to learn the frequency-specific spatial filters and reduce the number of trainable parameters. In the second block, a separable convolution is applied to learn how to individually summarize the feature maps and optimally combine them together. As the original EEGNet performed poorly when dealing with the epileptic EEG data \cite{lu2020staging}, we add two 1D CNN before the original layers to adequately extract the representation from each channel and concentrate on the important information. Meanwhile, we modify the size of convolution and average pooling kernels according to the sampling rate of our EEG signals. The structure of the modified EEGNet is shown in Fig. \ref{fig4}.

\subsubsection{Multi-scale CNN}
The structure of Multi-scale CNN can be divided into two parts: the temporal multi-scale part and the spatial multi-scale part. The dilated convolution is employed in both parts to broaden the receptive field and learn global information without increasing extra arguments. In each part, the convolutions with large kernels focus on information during a long term or over a large area of brain, while convolutions with small kernels capture the local information. Afterwards, an attention-based feature-weighted fusion method is taken to reduce the redundancy and fuse the features. The features are finally put into a four-layer CNN classifier.

\subsubsection{Transformer}
Transformer is a DL network characterized by the self-attention mechanism \cite{vaswani2017attention}\cite{liu2021swin}. Transformer models are generally composed of an encoder and a decoder. We only take advantage of the encoder part because the purpose of our task is classification. The Transformer-based model we used mainly consists of three modules: the input embedding module to diminish the input dimension and extract initial features, the positional encoding module to learn positional information which is then added to the embedding vector, and the attention module to explore associations between information at different locations and assign weights to them.

\subsection{Training and Testing}\label{Train}
\begin{figure}[htbp]
	\centerline{\includegraphics[width=0.45\textwidth]{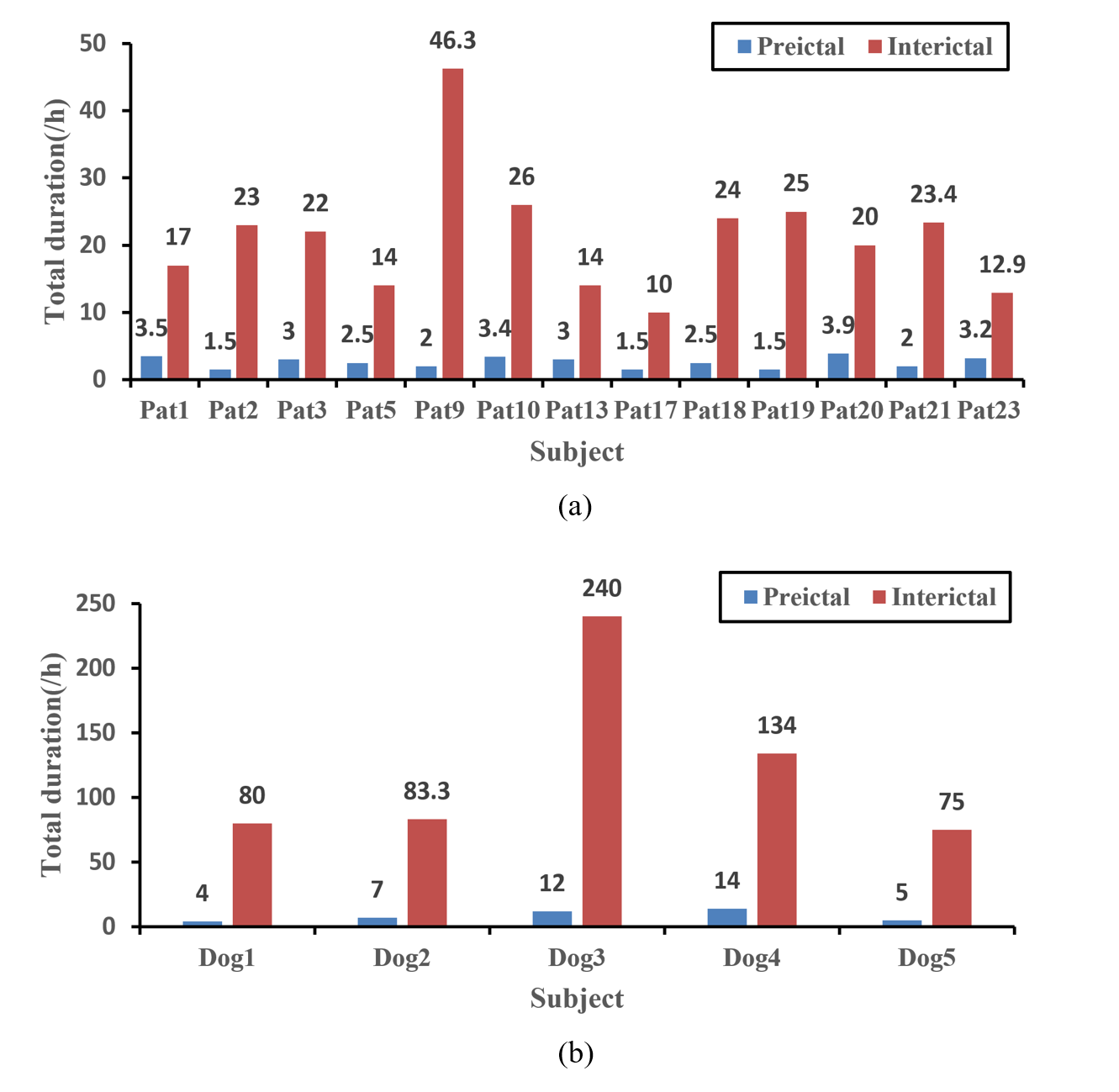}}
	\caption{(a) The total preictal and interictal duration of the CHB-MIT database; (b) The total preictal and interictal duration of the Kaggle database.}
	\label{fig5}
\end{figure}

In our work, we use the leave-one-out cross-validation strategy. Specifically, for a patient with $N$ seizures, we have $N$ parts of preictal data. So the whole interictal data is divided into $N$ equal parts accordingly. For every cross-validation, we take one part from the interictal data followed by one part from the preictal data as the test set while the rest of the interictal and preictal data serve as the training set. Due to the low-frequency of epileptic seizures, the total duration of the preictal data is significantly less than that of the interictal data. As we can see in Fig. \ref{fig5}, all subjects in the CHB-MIT and Kaggle database have an interictal-to-preictal ratio greater than 4:1. Some of them even exceed a ratio of 20:1. In order to avoid the adverse effects of imbalanced data on predictive models \cite{barandela2004imbalanced}, we employ four methods to solve the imbalanced problem:

\begin{itemize}
	\item Down-sampling \cite{khan2017focal}: Randomly picking samples from the interictal data until we get the same number as the preictal samples.
	\item Sliding windows \cite{truong2018convolutional}: Sliding windows of length $W$ at every step $S$ to create samples with overlapping parts.
	\item Recombination \cite{zhang2019epilepsy}: We first divide each preictal samples into three segments, and then randomly pick three segments to recombine new artificial samples.
	\item Diffusion model: We use the proposed DiffEEG to generate high-quality preictal samples.
\end{itemize}

\begin{figure*}[htbp]
	\centerline{\includegraphics[width=0.85\textwidth]{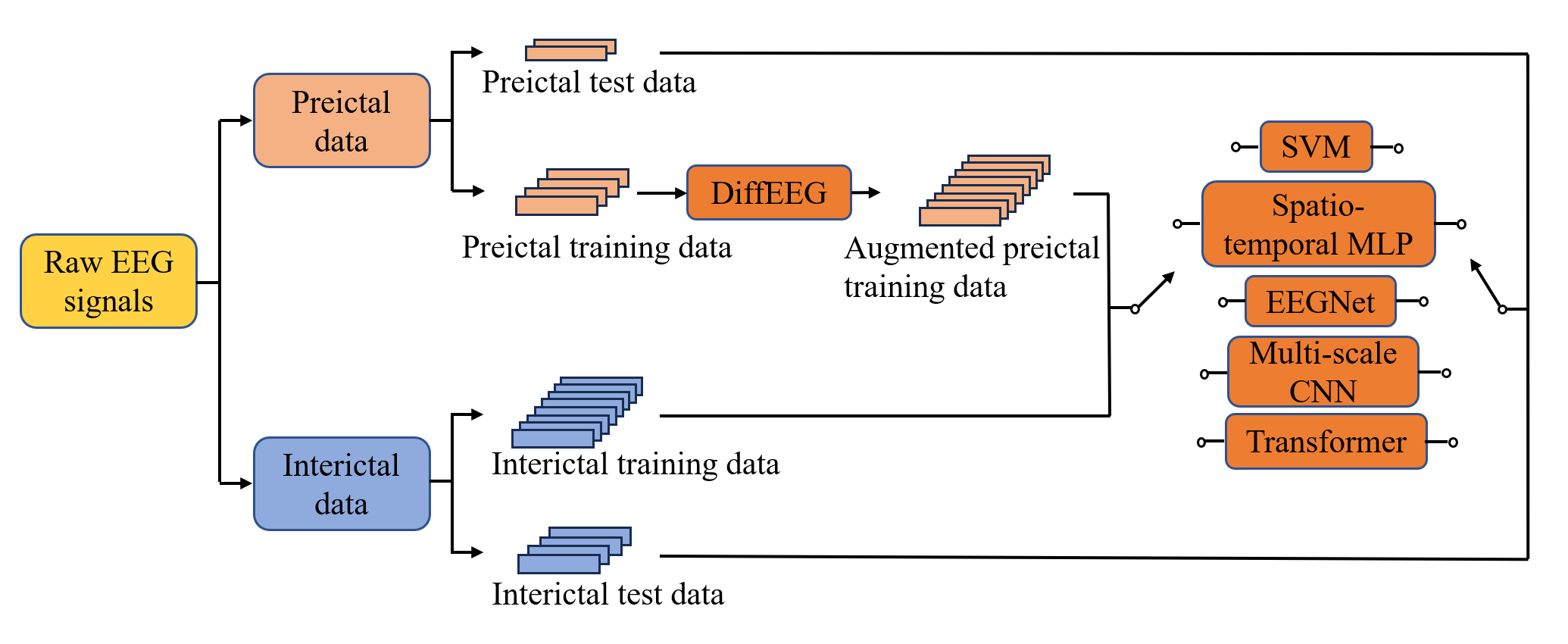}}
	\caption{A flowchart of the proposed method.}
	\label{fig6}
\end{figure*}

We randomly cut out EEG samples from the training set and send their STFT spectrograms to the trained model as condition. To further raise the diversity of produced samples, besides the randomly selected spectrograms, we also make use of the recombined spectrograms to each generate half of the data. We randomly pick out three preictal samples from the training set, divide the time dimension of their STFT spectrogram into three segments, and randomly choose one sub-spectrogram from each samples for recombination. The spectrogram is then put into the trained DiffEEG to generate EEG signal. After generating enough preictal samples to equal the interictal samples, we mix them randomly and select 25\% of the later samples in the training set for validation and the rest for training. Early-stop strategy is utilized to avoid overfitting. The training stops unless the sensitivity and specificity are both the best for five epochs and the best model is then returned. Fig. \ref{fig6} is a flowchart of the main process.

\subsection{Postprocessing}
To reduce false alarms, we use the $k$-of-$n$ method proposed in \cite{truong2018convolutional}. The alarm will be triggered only if at least $k$ out of the last $n$ predictions are judged as preictal. We set $k$ to 8 and $n$ to 10 in our experiments. In addition, a refractory period of 30 minutes is implemented to avoid frequent alarms in the short term.

\section{Experiments and Results}\label{Expe}
In this section, comparative experiments are conducted on the CHB-MIT and Kaggle database. The details of the two datasets and the evaluation metrics of our experiments are presented, followed by the experimental results and comparisons with several SOTA methods.  

\subsection{Dataset}
The CHB-MIT dataset records the sEEG data from 23 patients. It involves recordings for 844 hours with 163 seizure events, at a sampling rate of 256 Hz \cite{shoeb2009application}. As the electrode channels of different patients vary from 18 to 22, we select the 18 common electrodes for our research \cite{zhang2019epilepsy}. We follow the definitions of preictal and interictal period in \cite{truong2018convolutional}, and consider seizures with an interval of less than 15 minutes as a single seizure. Besides, We only focus on patients experiencing less than 10 seizures per day since it is not essential to perform the prediction task for patients who have an average seizure frequency of every 2 hours. Moreover, we select patients who experienced more than three seizures and had an interictal-to-preictal ratio greater than 2:1. This ensures that we have sufficient data to train the DiffEEG model and necessitates the generation of preictal samples. Based on these considerations, there are 13 patients that meet our requirements. The data information is listed in Table \ref{tab1}. 

\begin{table}[htbp]
	\caption{Subject information of the CHB-MIT sEEG database.}
	\begin{center}
		\resizebox{\linewidth}{!}{
			\begin{tabular}{cccc}
				\toprule
				Patient & Interictal times (h) & Preictal times (h) & No. of seizures \\
				\midrule
				Pat1 & 17 & 3.5 & 7 \\
				Pat2 & 23 & 1.5 & 3 \\
				Pat3 & 22 & 3 & 6 \\
				Pat5 & 14 & 2.5 & 5 \\
				Pat9 & 46.3 & 2 & 4 \\
				Pat10 & 26 & 3.4 & 7 \\
				Pat13 & 14 & 3 & 7 \\
				Pat17 & 10 & 1.5 & 3 \\
				Pat18 & 24 & 2.5 & 5 \\
				Pat19 & 25 & 1.5 & 3 \\
				Pat20 & 20 & 3.9 & 8 \\
				Pat21 & 23.4 & 2 & 4 \\
				Pat23 & 12.9 & 3.2 & 7 \\
				\midrule
				Total & 277.6 & 33.5 & 69 \\
				\bottomrule
			\end{tabular}
		}
		\label{tab1}
	\end{center}
\end{table}

The Kaggle dataset records the iEEG data from five dogs and two patients. It contains long-term recordings with 48 seizures \cite{brinkmann2016crowdsourcing}. For dogs, the electrode channels are 16 for Dog1-Dog4 and 15 for Dog5. The sampling rate is 400 Hz. For patients, the electrode channels are 15 for patient1 and 24 for patient2. The sampling rate is 5 kHz. As the high sampling rate could cause an increase in data dimension and model complexity, we don't test our method on these two patients. The data information of five dogs is listed in Table \ref{tab2}.

\begin{table}[htbp]
	\caption{Subject information of the Kaggle iEEG database.}
	\begin{center}
		\resizebox{\linewidth}{!}{
			\begin{tabular}{cccc}
				\toprule
				Subject & Interictal times (h) & Preictal times (h) & No. of seizures \\
				\midrule
				Dog1 & 80 & 4 & 4 \\
				Dog2 & 83.3 & 7 & 7 \\
				Dog3 & 240 & 12 & 12 \\
				Dog4 & 134 & 14 & 14 \\
				Dog5 & 75 & 5 & 5 \\
				\midrule
				Total & 612.3 & 42 & 42 \\
				\bottomrule
			\end{tabular}
		}
		\label{tab2}
	\end{center}
\end{table}

\subsection{Experiments and Evaluation Metrics}
Before the experiments, it is necessary to define the seizure prediction horizon (SPH) and the seizure occurrence period (SOP). We adopt the definitions proposed by Maiwald et al. \cite{maiwald2004comparison}. The SPH represents the time interval between the seizure alarm and the actual onset of seizure, while the SOP represents the period where seizures are predicted to happen. For a correct prediction, the seizure onset must fall within the SOP. If the system returns a preictal result but no seizure occurs during the SOP, then there is a false alarm. In our experiments, the SPH and SOP are set to 1 minute and 30 minutes for the CHB-MIT database, while for the Kaggle database, the SPH and SOP are set to 5 minutes and 60 minutes.

Seizure prediction experiments are carried on in an event-based way or segment-based way. According to \cite{zhang2023distilling}, the event-based prediction is more representative of practical situations and holds greater significance in real-world applications, so we take the event-based way. We then apply four evaluation metrics to estimate the prediction results. Sensitivity (Sens) is defined as the proportion of successfully predicted seizures to all seizures. False prediction rate (FPR) is defined as the number of false alarms within an hour. Area under the curve (AUC) can reflect the quality of classification performance, where a value of 0.5 represents a random classification and a value of 1 signifies a perfect classifier.

p-value can determine whether the prediction system demonstrates statistical superiority over a random predictor. According to \cite{gao2022general}, the sensitivity $S_{c}$ of chance predictor can be formulated as equation (\ref{equ:11}), where ${\lambda}_w$ represents the Poisson rate parameter, ${\tau}_{w_0}$ represents SPH and ${\tau}_w$ denotes the sum of SPH and SOP.
\begin{equation}\label{equ:11}
    S_{c} = 1 - exp(-{\lambda}_w{\tau}_w + (1 - e^{-{\lambda}_w{\tau}_{w_0}}))
\end{equation}
Under the assumption that the prediction system successfully predicts $n$ out of $N$ seizures, the random predictor can outperform the prediction system if it correctly predicts at least $n$ seizures. The p-value can be calculated as below at a significant level of 0.05:
\begin{equation}
    p\text{-}value = 1-\sum_{i=0}^{n-1}\binom{N}{i}S_{c}^i(1-S_{c})^{N-i},\ \text{for} \ \frac{n}{N}\geq S_{c}.
\end{equation}

\subsection{Results and Comparison}
\begin{figure}[htbp]
	\centerline{\includegraphics[width=0.45\textwidth]{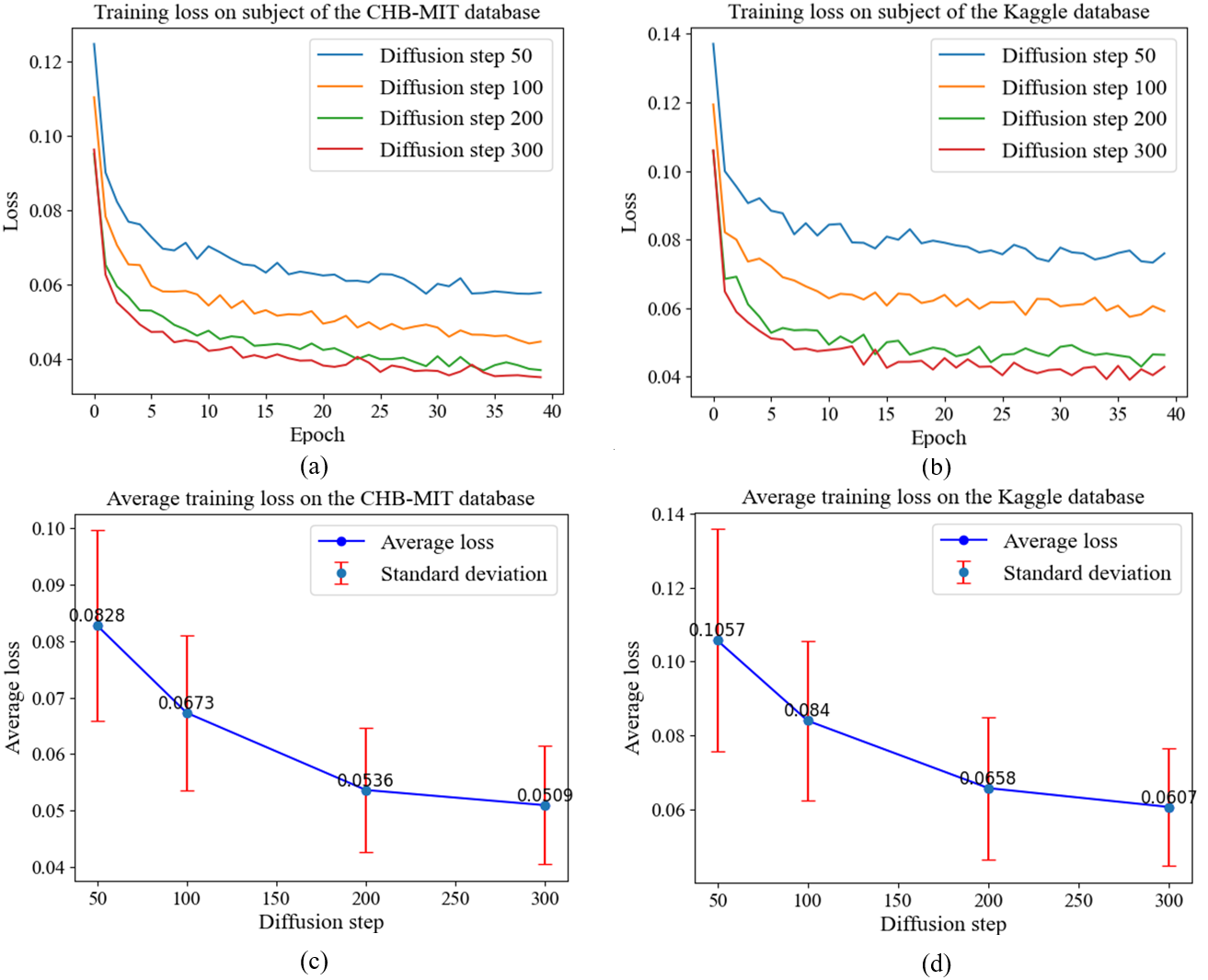}}
	\caption{The loss curves of different diffusion steps. (a)(b) are the loss curves against epochs of two subjects from the CHB-MIT and Kaggle dataset. (c)(d) are the average converged losses of all subjects from each dataset for different diffusion steps.}
	\label{fig7}
\end{figure}

To study the impact of different time steps on the generation quality of the DiffEEG model, we select four step values (50, 100, 200, 300) and train the model for 40 epochs respectively. Since the quality of synthetic samples is hard to quantitatively assess, we utilize the training loss as a proxy. A lower loss implies that the model has learned the data distribution and epileptic representation more comprehensively, resulting in higher quality of the generated samples. To observe the experimental results dynamically, we plot the loss curves against epochs for the four diffusion steps. In Fig. \ref{fig7}, we display the curves of two subjects from the CHB-MIT and Kaggle dataset respectively. In the meantime, the average converged loss and the standard deviation between losses of each subject are calculated for an overall comparison. From the curves, it can be observed that as time step increases from 50 to 200, the converged loss value significantly decreases. However, when time step is increased to 300, the decrease in the converged loss is not substantial. Therefore, we opt for 200 as the final value of diffusion step.

To demonstrate the effectiveness of using DiffEEG for DA, we conduct comparative experiments with classification networks using just the original data and the augmented data generated by DiffEEG respectively. To explore the superiority of DiffEEG in solving the imbalanced data problem, we conduct the contrast experiments with models using the down-sampling, sliding windows and recombination methods. In the meantime, we utilize five representative network frameworks: SVM, Spatio-temporal MLP, EEGNet, Multi-scale CNN, Transformer as classifiers and test the prediction performance on the CHB-MIT and Kaggle database, to validate the universality of our approach. The performance is concluded in Table \ref{tab3} $\sim$ Table \ref{tab12}. To make the comparative effectiveness more obvious, the best results among all methods are marked with bold.

\begin{table*}[htbp]
	\caption{Performance of SVM on CHB-MIT database.}
	\begin{center}
		\resizebox{\textwidth}{!}{
			\begin{tabular}{ccccccccccccccccccccc}
				\toprule
					Patient & \multicolumn{4}{c}{Original} & \multicolumn{4}{c}{Down-sampling} & \multicolumn{4}{c}{Sliding windows} & \multicolumn{4}{c}{Recombination} & \multicolumn{4}{c}{\textbf{DiffEEG}} \\
				\cline{2-21}
				& Sens (\%) & FPR/h & AUC & p-value & Sens (\%) & FPR/h & AUC & p-value & Sens (\%) & FPR/h & AUC & p-value & Sens (\%) & FPR/h & AUC & p-value & Sens (\%) & FPR/h & AUC & p-value \\
				\midrule
				Pat1 & 100.0 & 0.000 & 0.9669 & \textless0.001 & 100.0 & 0.000 & \textbf{0.9869} & \textless0.001 & 100.0 & 0.000 & 0.9711 & \textless0.001 & 100.0 & 0.000 & 0.9819 & \textless0.001 & \textbf{100.0} & \textbf{0.000} & 0.9821 & \textbf{\textless0.001} \\
				Pat2 & 66.7 & 0.000 & 0.7776 & \textbf{0.003} & 66.7 & 0.043 & 0.8104 & 0.009 & 66.7 & 0.000 & 0.8227 & 0.007 & 66.7 & 0.000 & 0.7773 & 0.004 & \textbf{66.7} & \textbf{0.000} & \textbf{0.8228} & 0.005 \\
				Pat3 & 66.7 & 0.136 & 0.7551 & \textless0.001 & 100.0 & 0.318 & 0.8301 & \textless0.001 & 100.0 & 0.136 & 0.8467 & \textless0.001 & 100.0 & 0.227 & 0.8444 & \textless0.001 & \textbf{100.0} & \textbf{0.091} & \textbf{0.8496} & \textbf{\textless0.001} \\
				Pat5 & 80.0 & 0.071 & 0.8071 & \textless0.001 & 80.0 & 0.071 & \textbf{0.8609} & 0.002 & 100.0 & 0.071 & 0.8471 & \textless0.001 & 80.0 & 0.071 & 0.8292 & 0.001 & \textbf{100.0} & \textbf{0.071} & 0.8332 & \textbf{\textless0.001} \\
				Pat9 & 25.0 & \textbf{0.022} & 0.6035 & \textbf{0.054} & 50.0 & 0.281 & 0.6954 & 0.075 & 50.0 & 0.259 & \textbf{0.7188} & 0.059 & 50.0 & 0.216 & 0.6729 & 0.068 & \textbf{50.0} & 0.259 & 0.6878 & 0.087 \\
				Pat10 & 71.4 & \textbf{0.346} & 0.6953 & 0.002 & 71.4 & 0.500 & 0.7119 & 0.009 & 57.1 & 0.462 & 0.7499 & 0.040 & 71.4 & 0.423 & 0.7288 & 0.004 & \textbf{85.7} & 0.462 & \textbf{0.7714} & \textbf{\textless0.001} \\
				Pat13 & 100.0 & \textbf{0.214} & 0.8867 & \textless0.001 & 100.0 & 0.357 & \textbf{0.9125} & \textless0.001 & 100.0 & 0.357 & 0.9057 & \textless0.001 & 100.0 & 0.357 & 0.8983 & \textless0.001 & \textbf{100.0} & 0.357 & 0.8899 & \textbf{\textless0.001} \\
				Pat17 & 66.7 & 0.200 & 0.7491 & 0.060 & 66.7 & 0.400 & 0.7375 & 0.090 & 66.7 & 0.200 & \textbf{0.7538} & \textbf{0.060} & 66.7 & 0.200 & 0.7480 & 0.070 & \textbf{66.7} & \textbf{0.200} & 0.7511 & 0.066 \\
				Pat18 & 60.0 & 0.000 & 0.7912 & 0.002 & 60.0 & 0.083 & 0.7967 & 0.005 & 60.0 & \textbf{0.000} & 0.7969 & 0.002 & 60.0 & 0.041 & 0.7845 & 0.002 & \textbf{80.0} & 0.083 & \textbf{0.8144} & \textbf{\textless0.001} \\
				Pat19 & 100.0 & 0.000 & 0.8000 & \textless0.001 & 100.0 & 0.120 & 0.8867 & \textless0.001 & 100.0 & 0.040 & 0.8885 & \textless0.001 & 100.0 & 0.040 & 0.8883 & \textless0.001 & \textbf{100.0} & \textbf{0.000} & \textbf{0.8990} & \textbf{\textless0.001} \\
				Pat20 & 100.0 & 0.050 & 0.9807 & \textless0.001 & 100.0 & 0.150 & 0.9748 & \textless0.001 & 100.0 & 0.050 & 0.9831 & \textless0.001 & 100.0 & 0.100 & \textbf{0.9850} & \textless0.001 & \textbf{100.0} & \textbf{0.000} & 0.9828 & \textbf{\textless0.001} \\
				Pat21 & 100.0 & 0.214 & 0.9297 & \textless0.001 & 100.0 & 0.299 & 0.9186 & 0.001 & 100.0 & \textbf{0.214} & 0.9401 & \textless0.001 & 100.0 & 0.256 & 0.9251 & \textless0.001 & \textbf{100.0} & 0.256 & \textbf{0.9457} & \textbf{\textless0.001} \\
				Pat23 & 100.0 & 0.077 & 0.9880 & \textless0.001 & 100.0 & 0.077 & 0.9849 & \textless0.001 & 100.0 & 0.077 & \textbf{0.9916} & \textless0.001 & 100.0 & 0.077 & 0.9903 & \textless0.001 & \textbf{100.0} & \textbf{0.077} & 0.9903 & \textbf{\textless0.001} \\
				\midrule
				Ave & 79.7 & \textbf{0.102} & 0.825 & - & 84.2 & 0.208 & 0.854 & - & 84.7 & 0.144 & 0.863 & - & 84.2 & 0.155 & 0.850 & - & \textbf{88.4} & 0.143 & \textbf{0.863} & \textbf{-} \\
				\bottomrule
			\end{tabular}
			\label{tab3}
		}
	\end{center}
\end{table*}

\begin{table*}[htbp]
	\caption{Performance of Spatio-temporal MLP on CHB-MIT database.}
	\begin{center}
		\resizebox{\textwidth}{!}{
			\begin{tabular}{ccccccccccccccccccccc}
				\toprule
					Patient & \multicolumn{4}{c}{Original} & \multicolumn{4}{c}{Down-sampling} & \multicolumn{4}{c}{Sliding windows} & \multicolumn{4}{c}{Recombination} & \multicolumn{4}{c}{\textbf{DiffEEG}} \\
				\cline{2-21}
				& Sens (\%) & FPR/h & AUC & p-value & Sens (\%) & FPR/h & AUC & p-value & Sens (\%) & FPR/h & AUC & p-value & Sens (\%) & FPR/h & AUC & p-value & Sens (\%) & FPR/h & AUC & p-value \\
				\midrule
				Pat1 & 100.0 & 0.000 & 0.9663 & \textless0.001 & 100.0 & 0.000 & 0.9806 & \textless0.001 & 100.0 & 0.000 & 0.9711 & \textless0.001 & 100.0 & 0.000 & 0.9834 & \textless0.001 & \textbf{100.0} & \textbf{0.000} & \textbf{0.9935} & \textbf{\textless0.001} \\
				Pat2 & 66.7 & 0.000 & 0.6302 & \textless0.001 & 100.0 & 0.043 & 0.7336 & \textless0.001 & 100.0 & 0.000 & 0.7755 & \textless0.001 & 100.0 & 0.000 & 0.8144 & \textless0.001 & \textbf{100.0} & \textbf{0.000} & \textbf{0.8537} & \textbf{\textless0.001} \\
				Pat3 & 100.0 & 0.045 & 0.8656 & \textless0.001 & 83.3 & 0.091 & 0.8451 & \textless0.001 & 100.0 & 0.045 & 0.8966 & \textless0.001 & 100.0 & 0.091 & 0.8611 & \textless0.001 & \textbf{100.0} & \textbf{0.045} & \textbf{0.9171} & \textbf{\textless0.001} \\
				Pat5 & 80.0 & 0.000 & 0.8424 & 0.002 & 100.0 & 0.285 & 0.8380 & \textless0.001 & 100.0 & 0.071 & 0.9082 & \textless0.001 & 100.0 & 0.142 & 0.8979 & 0.001 & \textbf{100.0} & \textbf{0.000} & \textbf{0.9458} & \textbf{\textless0.001} \\
				Pat9 & 50.0 & \textbf{0.022} & 0.6085 & \textless0.001 & 75.0 & 0.216 & 0.6951 & \textless0.001 & 75.0 & 0.064 & 0.6262 & \textless0.001 & 75.0 & 0.064 & 0.7101 & \textless0.001 & \textbf{75.0} & 0.043 & \textbf{0.7651} & \textbf{\textless0.001} \\
				Pat10 & 57.1 & 0.192 & 0.7042 & 0.005 & 71.4 & 0.269 & 0.7630 & 0.192 & 71.4 & 0.038 & 0.8190 & 0.032 & 57.1 & 0.153 & 0.7436 & 0.084 & \textbf{85.7} & \textbf{0.038} & \textbf{0.8435} & \textbf{0.001} \\
				Pat13 & 100.0 & 0.285 & 0.8578 & \textless0.001 & 100.0 & 0.285 & 0.8903 & \textless0.001 & 100.0 & 0.285 & 0.8683 & \textless0.001 & 100.0 & 0.285 & 0.8913 & \textless0.001 & \textbf{100.0} & \textbf{0.214} & \textbf{0.9086} & \textbf{\textless0.001} \\
				Pat17 & 66.7 & 0.300 & 0.7245 & 0.325 & 100.0 & 0.400 & 0.8044 & 0.199 & 100.0 & 0.400 & 0.7883 & 0.114 & 100.0 & 0.300 & 0.8082 & 0.057 & \textbf{100.0} & \textbf{0.200} & \textbf{0.8740} & \textbf{0.002} \\
				Pat18 & 80.0 & 0.042 & 0.8897 & 0.001 & 80.0 & 0.083 & 0.8012 & 0.005 & 80.0 & 0.000 & 0.8933 & \textless0.001 & 80.0 & 0.042 & 0.8722 & 0.003 & \textbf{80.0} & \textbf{0.000} & \textbf{0.9071} & \textbf{\textless0.001} \\
				Pat19 & 100.0 & 0.000 & \textbf{0.9570} & \textless0.001 & 66.7 & 0.000 & 0.6786 & \textless0.001 & 66.7 & 0.000 & 0.8013 & 0.008 & 66.7 & 0.000 & 0.7950 & 0.008 & \textbf{100.0} & \textbf{0.000} & 0.9488 & \textbf{\textless0.001} \\
				Pat20 & 87.5 & 0.050 & 0.8949 & \textless0.001 & 100.0 & 0.150 & 0.9380 & \textless0.001 & 100.0 & 0.050 & 0.9542 & \textless0.001 & 100.0 & 0.050 & 0.9454 & \textless0.001 & \textbf{100.0} & \textbf{0.050} & \textbf{0.9792} & \textbf{\textless0.001} \\
				Pat21 & 50.0 & 0.214 & 0.6013 & \textless0.001 & 100.0 & 0.427 & 0.7237 & 0.011 & 100.0 & 0.214 & 0.7182 & \textless0.001 & 100.0 & 0.214 & 0.7683 & \textless0.001 & \textbf{100.0} & \textbf{0.214} & \textbf{0.8198} & \textbf{\textless0.001} \\
				Pat23 & 100.0 & 0.000 & 0.9930 & \textless0.001 & 100.0	& 0.000	& 0.9867 & \textless0.001 & 100.0 & 0.000 & 0.9795 & \textless0.001 & 100.0 & 0.000 & 0.9923 & \textless0.001 & \textbf{100.0} & \textbf{0.000} & \textbf{0.9969} & \textbf{\textless0.001} \\
				\midrule
				Ave & 79.8 & 0.089 & 0.810 & - & 90.5 & 0.173 & 0.821 & - & 91.8 & 0.090 & 0.846 & - & 90.7 & 0.103 & 0.853 & - & \textbf{95.4} & \textbf{0.062} & \textbf{0.904} & \textbf{-} \\
				\bottomrule
			\end{tabular}
			\label{tab4}
		}
	\end{center}
\end{table*}

\begin{table*}[htbp]
	\caption{Performance of EEGNet on CHB-MIT database.}
	\begin{center}
		\resizebox{\textwidth}{!}{
			\begin{tabular}{ccccccccccccccccccccc}
				\toprule
					Patient & \multicolumn{4}{c}{Original} & \multicolumn{4}{c}{Down-sampling} & \multicolumn{4}{c}{Sliding windows} & \multicolumn{4}{c}{Recombination} & \multicolumn{4}{c}{\textbf{DiffEEG}} \\
				\cline{2-21}
				& Sens (\%) & FPR/h & AUC & p-value & Sens (\%) & FPR/h & AUC & p-value & Sens (\%) & FPR/h & AUC & p-value & Sens (\%) & FPR/h & AUC & p-value & Sens (\%) & FPR/h & AUC & p-value \\
				\midrule
				Pat1 & 100.0 & 0.000 & 0.9702 & \textless0.001 & 100.0 & 0.000 & 0.9671 & \textless0.001 & 100.0 & 0.000 & 0.9819 & \textless0.001 & 100.0 & 0.000 & 0.9785 & \textless0.001 & \textbf{100.0} & \textbf{0.000} & \textbf{0.9857} & \textbf{\textless0.001} \\
				Pat2 & 0.0 & 0.000 & 0.5068 & 1.000 & 33.3 & 0.087 & 0.5664 & 0.121 & 66.7 & 0.000 & 0.6186 & 0.036 & 33.3 & 0.000 & 0.5690 & 0.091 & \textbf{66.7} & \textbf{0.000} & \textbf{0.7150} & \textbf{0.037} \\
				Pat3 & 66.7 & 0.000 & 0.7938 & \textless0.001 & 66.7 & 0.091 & 0.7796 & \textless0.001 & 83.3 & 0.000 & 0.8638 & \textless0.001 & 66.7 & 0.000 & 0.8566 & 0.001 & \textbf{83.3} & \textbf{0.000} & \textbf{0.8783} & \textbf{\textless0.001} \\
				Pat5 & 80.0 & 0.000 & 0.8621 & \textless0.001 & 100.0 & 0.357 & 0.8056 & \textless0.001 & 100.0 & 0.000 & 0.9489 & \textless0.001 & 100.0 & 0.000 & 0.9131 & \textless0.001 & \textbf{100.0} & \textbf{0.000} & \textbf{0.9639} & \textbf{\textless0.001} \\
				Pat9 & 0.0 & 0.000 & 0.5000 & 1.000 & 75.0 & 0.367 & 0.6563 & 0.009 & 50.0 & \textbf{0.000} & 0.7140 & 0.197 & 75.0 & 0.043 & 0.7425 & 0.005 & \textbf{75.0} & 0.086 & \textbf{0.7712} & \textbf{\textless0.001} \\
				Pat10 & 57.1 & 0.077 & 0.6878 & \textless0.001 & 85.7 & 0.346 & 0.6906 & 0.012 & 85.7 & \textbf{0.038} & 0.8288 & \textless0.001 & 85.7 & 0.115 & 0.8062 & \textless0.001 & \textbf{85.7} & 0.077 & \textbf{0.8610} & \textbf{\textless0.001} \\
				Pat13 & 100.0 & 0.286 & 0.8799 & \textless0.001 & 100.0 & 0.286 & 0.8783 & \textless0.001 & 100.0 & 0.286 & 0.9060 & \textless0.001 & 100.0 & 0.286 & 0.9046 & \textless0.001 & \textbf{100.0} & \textbf{0.286} & \textbf{0.9278} & \textbf{\textless0.001} \\
				Pat17 & 66.7 & 0.200 & 0.7361 & 0.037 & 100.0 & 0.200 & 0.7793 & \textbf{0.001} & 66.7 & 0.300 & 0.7538 & 0.052 & 66.7 & 0.300 & 0.7688 & 0.075 & \textbf{100.0} & \textbf{0.200} & \textbf{0.9433} & 0.044 \\
				Pat18 & 40.0 & 0.042 & 0.6685 & 0.020 & 80.0 & 0.208 & 0.6976 & \textless0.001 & 80.0 & \textbf{0.000} & 0.8467 & \textless0.001 & 80.0 & 0.042 & 0.8317 & \textless0.001 & \textbf{80.0} & 0.042 & \textbf{0.8520} & \textbf{\textless0.001} \\
				Pat19 & 100.0 & 0.000 & 0.8264 & 0.002 & 66.7 & 0.040 & 0.7043 & 0.004 & 100.0 & 0.000 & 0.9114 & \textless0.001 & 100.0 & 0.000 & 0.8969 & \textless0.001 & \textbf{100.0} & \textbf{0.000} & \textbf{0.9719} & \textbf{\textless0.001} \\
				Pat20 & 100.0 & 0.050 & 0.9430 & \textless0.001 & 100.0 & 0.150 & 0.9429 & \textless0.001 & 100.0 & 0.100 & 0.9553 & \textless0.001 & 100.0 & 0.100 & 0.9585 & \textless0.001 & \textbf{100.0} & \textbf{0.050} & \textbf{0.9761} & \textbf{\textless0.001} \\
				Pat21 & 100.0 & \textbf{0.128} & 0.7296 & \textless0.001 & 100.0 & 0.299 & 0.7847 & 0.001 & 100.0 & 0.256 & 0.8484 & \textless0.001 & 100.0 & 0.256 & 0.8182 & \textless0.001 & \textbf{100.0} & 0.214 & \textbf{0.8746} & \textbf{\textless0.001} \\
				Pat23 & 100.0 & 0.000 & 0.9862 & \textless0.001 & 100.0 & 0.077 & 0.9673 & \textless0.001 & 100.0 & 0.000 & 0.9977 & \textless0.001 & 100.0 & 0.000 & 0.9945 & \textless0.001 & \textbf{100.0} & \textbf{0.000} & \textbf{0.9984} & \textbf{\textless0.001} \\
				\midrule
				Ave & 70.0 & \textbf{0.060} & 0.776 & - & 85.2 & 0.193 & 0.786 & - & 87.1 & 0.075 & 0.860 & - & 85.2 & 0.088 & 0.849 & - & \textbf{91.6} & 0.073 & \textbf{0.901} & \textbf{-} \\
				\bottomrule
			\end{tabular}
			\label{tab5}
		}
	\end{center}
\end{table*}

\begin{table*}[htbp]
	\caption{Performance of Multi-scale CNN on CHB-MIT database.}
	\begin{center}
		\resizebox{\textwidth}{!}{
			\begin{tabular}{ccccccccccccccccccccc}
				\toprule
				Patient & \multicolumn{4}{c}{Original} & \multicolumn{4}{c}{Down-sampling} & \multicolumn{4}{c}{Sliding windows} & \multicolumn{4}{c}{Recombination} & \multicolumn{4}{c}{\textbf{DiffEEG}} \\
				\cline{2-21}
				& Sens (\%) & FPR/h & AUC & p-value & Sens (\%) & FPR/h & AUC & p-value & Sens (\%) & FPR/h & AUC & p-value & Sens (\%) & FPR/h & AUC & p-value & Sens (\%) & FPR/h & AUC & p-value \\
				\midrule
				Pat1 & 100.0 & 0.000 & 0.9721 & \textless0.001 & 100.0 & 0.000 & 0.9799 & \textless0.001 & 100.0 & 0.000 & 0.9812 & \textless0.001 & 100.0 & 0.000 & 0.9927 & \textless0.001 & \textbf{100.0} & \textbf{0.000} & \textbf{0.9989} & \textbf{\textless0.001} \\
				Pat2 & 66.7 & 0.000 & 0.6331 & \textless0.001 & 66.7 & 0.087 & 0.6783 & \textless0.001 & 100.0 & 0.000 & 0.6837 & \textless0.001 & 66.7 & \textbf{0.000} & 0.7462 & \textless0.001 & \textbf{100.0} & 0.044 & \textbf{0.8884} & \textbf{\textless0.001} \\
				Pat3 & 100.0 & 0.045 & 0.8944 & \textless0.001 & 100.0 & 0.045 & 0.8735 & \textless0.001 & 100.0 & 0.045 & 0.9062 & \textless0.001 & 100.0 & 0.045 & 0.9319 & \textless0.001 & \textbf{100.0} & \textbf{0.000} & \textbf{0.9596} & \textbf{\textless0.001} \\
				Pat5 & 80.0 & 0.000 & 0.8917 & 0.002 & 100.0 & 0.214 & 0.8896 & 0.003 & 80.0 & 0.000 & 0.9228 & 0.002 & 100.0 & 0.000 & 0.9509 & \textless0.001 & \textbf{100.0} & \textbf{0.000} & \textbf{0.9961} & \textbf{\textless0.001} \\
				Pat9 & 50.0 & 0.022 & 0.7075 & 0.004 & 75.0 & 0.172 & 0.6552 & 0.106 & 75.0 & 0.064 & 0.7402 & \textless0.001 & 75.0 & 0.043 & 0.7722 & \textless0.001 & \textbf{75.0} & \textbf{0.000} & \textbf{0.8244} & \textbf{\textless0.001} \\
				Pat10 & 85.7 & 0.077 & 0.7294 & \textless0.001 & 85.7 & 0.307 & 0.7447 & \textless0.001 & 85.7 & 0.115 & 0.7529 & \textless0.001 & 85.7 & 0.192 & 0.8098 & \textless0.001 & \textbf{85.7} & \textbf{0.077} & \textbf{0.8795} & \textbf{\textless0.001} \\
				Pat13 & 71.4 & 0.500 & 0.7863 & 0.154 & 85.7 & \textbf{0.142} & 0.8352 & 0.142 & 85.7 & 0.500 & 0.7868 & 0.147 & 100.0 & 0.428 & 0.8730 & 0.003 & \textbf{100.0} & 0.214 & \textbf{0.9469} & \textbf{\textless0.001} \\
				Pat17 & 66.7 & 0.200 & 0.7123 & 0.193 & 100.0 & 0.200 & 0.8618 & 0.014 & 100.0 & 0.300 & \textbf{0.8804} & 0.042 & 100.0 & 0.300 & 0.7917 & 0.012 & \textbf{100.0} & \textbf{0.100} & 0.8681 & \textbf{\textless0.001} \\
				Pat18 & 80.0 & 0.000 & 0.8923 & \textless0.001 & 80.0 & 0.083 & 0.8760 & \textless0.001 & 80.0 & 0.041 & 0.8854 & \textless0.001 & 80.0 & 0.000 & 0.8920 & \textless0.001 & \textbf{80.0} & \textbf{0.000} & \textbf{0.8987} & \textbf{\textless0.001} \\
				Pat19 & 100.0 & 0.040 & 0.8584 & \textless0.001 & 100.0 & 0.200 & 0.7919 & 0.003 & 100.0 & 0.080 & 0.9069 & \textless0.001 & 100.0 & 0.000 & 0.9746 & \textless0.001 & \textbf{100.0} & \textbf{0.000} & \textbf{0.9898} & \textbf{\textless0.001} \\
				Pat20 & 100.0 & 0.200 & 0.9398 & \textless0.001 & 100.0 & 0.300 & 0.9337 & \textless0.001 & 100.0 & 0.150 & 0.9636 & \textless0.001 & 100.0 & 0.100 & 0.9723 & \textless0.001 & \textbf{100.0} & \textbf{0.050} & \textbf{0.9853} & \textbf{\textless0.001} \\
				Pat21 & 100.0 & 0.256 & 0.7291 & 0.014 & 100.0 & 0.427 & 0.7548 & 0.198 & 100.0 & 0.256 & 0.8448 & \textless0.001 & 100.0 & 0.300 & \textbf{0.9052} & \textless0.001 & \textbf{100.0} & \textbf{0.182} & 0.8967 & \textbf{\textless0.001} \\
				Pat23 & 100.0 & 0.000 & 0.9819 & \textless0.001 & 100.0 & 0.077 & 0.9800 & \textless0.001 & 100.0 & 0.077 & 0.9890 & \textless0.001 & 100.0 & 0.000 & \textbf{0.9923} & \textless0.001 & \textbf{100.0} & \textbf{0.000} & 0.9872 & \textbf{\textless0.001} \\
				\midrule
				Ave & 84.7 & 0.103 & 0.825 & - & 91.8 & 0.173 & 0.835 & - & 92.8 & 0.125 & 0.865 & - & 92.9 & 0.108 & 0.893 & - & \textbf{95.4} & \textbf{0.051} & \textbf{0.932} & \textbf{-} \\
				\bottomrule
			\end{tabular}
			\label{tab6}
		}
	\end{center}
\end{table*}

\begin{table*}[htbp]
	\caption{Performance of Transformer on CHB-MIT database.}
	\begin{center}
		\resizebox{\textwidth}{!}{
			\begin{tabular}{ccccccccccccccccccccc}
				\toprule
				Patient & \multicolumn{4}{c}{Original} & \multicolumn{4}{c}{Down-sampling} & \multicolumn{4}{c}{Sliding windows} & \multicolumn{4}{c}{Recombination} & \multicolumn{4}{c}{\textbf{DiffEEG}} \\
				\cline{2-21}
				& Sens (\%) & FPR/h & AUC & p-value & Sens (\%) & FPR/h & AUC & p-value & Sens (\%) & FPR/h & AUC & p-value & Sens (\%) & FPR/h & AUC & p-value & Sens (\%) & FPR/h & AUC & p-value \\
				\midrule
				Pat1 & 100.0 & 0.059 & 0.9542 & \textless0.001 & 100.0 & 0.000 & 0.9861 & \textless0.001 & 100.0 & 0.000 & 0.9899 & \textless0.001 & 100.0 & 0.059 & 0.9854 & \textless0.001 & \textbf{100.0} & \textbf{0.000} & \textbf{0.9954} & \textbf{\textless0.001} \\
				Pat2 & 33.3 & 0.000 & 0.6295 & 0.011 & 100.0 & 0.087 & 0.7642 & 0.002 & 100.0 & 0.000 & 0.8321 & \textless0.001 & 100.0 & 0.000 & 0.8205 & \textless0.001 & \textbf{100.0} & \textbf{0.000} & \textbf{0.9096} & \textbf{\textless0.001} \\
				Pat3 & 66.7 & \textbf{0.000} & 0.7652 & \textless0.001 & 66.7 & 0.090 & 0.7776 & \textless0.001 & 83.3 & 0.045 & 0.8835 & \textless0.001 & 100.0 & 0.136 & 0.8873 & \textless0.001 & \textbf{100.0} & 0.045 & \textbf{0.9225} & \textbf{\textless0.001} \\
				Pat5 & 60.0 & \textbf{0.000} & 0.7164 & 0.105 & 100.0 & 0.142 & 0.7840 & 0.002 & 100.0 & 0.071 & 0.8646 & \textless0.001 & 100.0 & 0.428 & 0.8175 & \textless0.001 & \textbf{100.0} & 0.142 & \textbf{0.9344} & \textbf{\textless0.001} \\
				Pat9 & 0.0 & \textbf{0.022} & 0.5333 & 1.000 & 50.0 & 0.086 & 0.6327 & 0.141 & 75.0 & 0.064 & 0.7738 & \textless0.001 & 75.0 & 0.108 & 0.7869 & \textless0.001 & \textbf{75.0} & 0.043 & \textbf{0.8384} & \textbf{\textless0.001} \\
				Pat10 & 28.6 & 0.115 & 0.5620 & 0.122 & 85.7 & 0.307 & 0.7130 & 0.197 & 85.7 & 0.192 & 0.7795 & 0.005 & 85.7 & 0.038 & 0.7833 & 0.011 & \textbf{85.7} & \textbf{0.077} & \textbf{0.8556} & \textbf{\textless0.001} \\
				Pat13 & 100.0 & 0.285 & 0.8986 & \textless0.001 & 100.0 & 0.285 & 0.9204 & \textless0.001 & 100.0 & 0.357 & 0.9303 & \textless0.001 & 100.0 & 0.214 & 0.9353 & \textless0.001 & \textbf{100.0} & \textbf{0.214} & \textbf{0.9589} & \textbf{\textless0.001} \\
				Pat17 & 66.7 & 0.100 & 0.8093 & 0.117 & 100.0 & 0.100 & 0.8946 & 0.132 & 100.0 & 0.100 & 0.8750 & 0.020 & 100.0 & 0.100 & 0.9017 & 0.015 & \textbf{100.0} & \textbf{0.100} & \textbf{0.9421} & \textbf{\textless0.001} \\
				Pat18 & 40.0 & 0.000 & 0.6735 & 0.042 & 80.0 & 0.083 & 0.7412 & 0.019 & 80.0 & 0.042 & 0.8069 & \textless0.001 & 80.0 & 0.000 & 0.7849 & \textless0.001 & \textbf{80.0} & \textbf{0.000} & \textbf{0.8331} & \textbf{\textless0.001} \\
				Pat19 & 0.0 & 0.000 & 0.5628 & 1.000 & 66.7 & 0.400 & 0.7739 & 0.110 & 66.7 & 0.080 & 0.7869 & 0.004 & 66.7 & 0.000 & 0.7857 & 0.035 & \textbf{100.0} & \textbf{0.040} & \textbf{0.9061} & \textbf{\textless0.001} \\
				Pat20 & 100.0 & 0.050 & 0.9074 & \textless0.001 & 100.0 & 0.150 & 0.9435 & \textless0.001 & 100.0 & 0.100 & 0.9631 & \textless0.001 & 100.0 & 0.100 & 0.9604 & \textless0.001 & \textbf{100.0} & \textbf{0.050} & \textbf{0.9828} & \textbf{\textless0.001} \\
				Pat21 & 75.0 & 0.085 & 0.7223 & 0.004 & 100.0 & 0.427 & 0.7976 & \textless0.001 & 100.0 & 0.128 & 0.8654 & \textless0.001 & 100.0 & 0.085 & 0.8819 & \textless0.001 & \textbf{100.0} & \textbf{0.085} & \textbf{0.9228} & \textbf{\textless0.001} \\
				Pat23 & 100.0 & 0.000 & 0.9760 & \textless0.001 & 100.0 & 0.000 & 0.9808 & \textless0.001 & 100.0 & 0.000 & 0.9910 & \textless0.001 & 100.0 & 0.000 & 0.9916 & \textless0.001 & \textbf{100.0} & \textbf{0.000} & \textbf{0.9941} & \textbf{\textless0.001} \\
				\midrule
				Ave & 59.3 & \textbf{0.055} & 0.747 & - & 88.4 & 0.166 & 0.824 & - & 91.6 & 0.091 & 0.872 & - & 92.9 & 0.098 & 0.871 & - & \textbf{95.4} & 0.061 & \textbf{0.923} & \textbf{-} \\
				\bottomrule
			\end{tabular}
			\label{tab7}
		}
	\end{center}
\end{table*}

\begin{table*}[htbp]
	\caption{Performance of SVM on Kaggle database.}
	\begin{center}
		\resizebox{\textwidth}{!}{
			\begin{tabular}{ccccccccccccccccccccc}
				\toprule
				Patient & \multicolumn{4}{c}{Original} & \multicolumn{4}{c}{Down-sampling} & \multicolumn{4}{c}{Sliding windows} & \multicolumn{4}{c}{Recombination} & \multicolumn{4}{c}{\textbf{DiffEEG}} \\
				\cline{2-21}
				& Sens (\%) & FPR/h & AUC & p-value & Sens (\%) & FPR/h & AUC & p-value & Sens (\%) & FPR/h & AUC & p-value & Sens (\%) & FPR/h & AUC & p-value & Sens (\%) & FPR/h & AUC & p-value \\
				\midrule
				Dog1 & 0.0 & \textbf{0.025} & 0.4989 & 1.000 & 50.0 & 0.213 & 0.5769 & 0.142 & 50.0 & 0.300 & 0.5779 & 0.127 & 50.0 & 0.263 & 0.6264 & 0.110 & \textbf{50.0} & 0.238 & \textbf{0.6282} & \textbf{0.088} \\
				Dog2 & 85.7 & \textbf{0.036} & 0.8334 & \textless0.001 & 100.0 & 0.072 & 0.9310 & \textless0.001 & 85.7 & 0.060 & 0.9110 & \textless0.001 & 100.0 & 0.072 & 0.9338 & \textless0.001 & \textbf{100.0} & 0.060 & \textbf{0.9545} & \textbf{\textless0.001} \\
				Dog3 & 66.7 & \textbf{0.017} & 0.6961 & \textless0.001 & 75.0 & 0.250 & 0.7724 & \textless0.001 & 83.3 & 0.242 & 0.7770 & \textless0.001 & 83.3 & 0.242 & 0.7760 & \textless0.001 & \textbf{83.3} & 0.183 & \textbf{0.7841} & \textbf{\textless0.001} \\
				Dog4 & 64.3 & \textbf{0.052} & 0.6286 & \textless0.001 & 92.9 & 0.269 & \textbf{0.7589} & \textless0.001 & 85.7 & 0.276 & 0.7383 & \textless0.001 & 92.9 & 0.276 & 0.7376 & \textless0.001 & \textbf{92.9} & 0.224 & 0.7097 & \textbf{\textless0.001} \\
				Dog5 & 60.0 & \textbf{0.040} & 0.7820 & 0.001 & 80.0 & 0.093 & 0.7994 & \textless0.001 & 80.0 & 0.053 & 0.7942 & \textless0.001 & 80.0 & 0.053 & 0.7929 & \textless0.001 & \textbf{100.0} & 0.053 & \textbf{0.8056} & \textbf{\textless0.001} \\
				\midrule
				Ave & 55.3 & \textbf{0.034} & 0.688 & - & 79.6 & 0.179 & 0.768 & - & 76.9 & 0.186 & 0.76 & - & 81.2 & 0.181 & 0.773 & - & \textbf{85.2} & 0.152 & \textbf{0.776} & \textbf{-} \\
				\bottomrule
			\end{tabular}
			\label{tab8}
		}
	\end{center}
\end{table*}

\begin{table*}[htbp]
	\caption{Performance of Spatio-temporal MLP on Kaggle database.}
	\begin{center}
		\resizebox{\textwidth}{!}{
			\begin{tabular}{ccccccccccccccccccccc}
				\toprule
				Patient & \multicolumn{4}{c}{Original} & \multicolumn{4}{c}{Down-sampling} & \multicolumn{4}{c}{Sliding windows} & \multicolumn{4}{c}{Recombination} & \multicolumn{4}{c}{\textbf{DiffEEG}} \\
				\cline{2-21}
				& Sens (\%) & FPR/h & AUC & p-value & Sens (\%) & FPR/h & AUC & p-value & Sens (\%) & FPR/h & AUC & p-value & Sens (\%) & FPR/h & AUC & p-value & Sens (\%) & FPR/h & AUC & p-value \\
				\midrule
				Dog1 & 0.0 & \textbf{0.000} & 0.5000 & 1.000 & 50.0 & 0.262 & 0.5491 & 0.167 & 50.0 & 0.200 & 0.5907 & 0.120 & 50.0 & 0.225 & 0.5638 & 0.080 & \textbf{75.0} & 0.250 & \textbf{0.6213} & \textbf{0.042} \\
				Dog2 & 57.1 & \textbf{0.048} & 0.6102 & \textless0.001 & 100.0 & 0.264 & 0.7866 & \textless0.001 & 100.0 & 0.228 & 0.8311 & \textless0.001 & 100.0 & 0.288 & 0.8229 & \textless0.001 & \textbf{100.0} & 0.216 & \textbf{0.8607} & \textbf{\textless0.001} \\
				Dog3 & 33.3 & \textbf{0.025} & 0.5966 & \textless0.001 & 83.3 & 0.179 & 0.7196 & \textless0.001 & 83.3 & 0.170 & 0.7215 & \textless0.001 & 83.3 & 0.158 & 0.7492 & \textless0.001 & \textbf{91.7} & 0.154 & \textbf{0.7855} & \textbf{\textless0.001} \\
				Dog4 & 14.3 & \textbf{0.075} & 0.5333 & 0.031 & 85.7 & 0.358 & 0.6632 & \textless0.001 & 85.7 & 0.373 & 0.6652 & \textless0.001 & 92.9 & 0.380 & 0.6719 & \textless0.001 & \textbf{92.9} & 0.298 & \textbf{0.7136} & \textbf{\textless0.001} \\
				Dog5 & 60.0 & \textbf{0.027} & 0.6584 & 0.007 & 100.0 & 0.173 & 0.8153 & \textless0.001 & 100.0 & 0.133 & 0.8313 & \textless0.001 & 100.0 & 0.133 & 0.8521 & \textless0.001 & \textbf{100.0} & 0.093 & \textbf{0.8641} & \textbf{\textless0.001} \\
				\midrule
				Ave & 32.9 & \textbf{0.035} & 0.580 & - & 83.8 & 0.247 & 0.707 & - & 83.8 & 0.221 & 0.728 & - & 85.2 & 0.237 & 0.732 & - & \textbf{91.9} & 0.202 & \textbf{0.769} & \textbf{-} \\
				\bottomrule
			\end{tabular}
			\label{tab9}
		}
	\end{center}
\end{table*}

\begin{table*}[htbp]
	\caption{Performance of EEGNet on Kaggle database.}
	\begin{center}
		\resizebox{\textwidth}{!}{
			\begin{tabular}{ccccccccccccccccccccc}
				\toprule
				Patient & \multicolumn{4}{c}{Original} & \multicolumn{4}{c}{Down-sampling} & \multicolumn{4}{c}{Sliding windows} & \multicolumn{4}{c}{Recombination} & \multicolumn{4}{c}{\textbf{DiffEEG}} \\
				\cline{2-21}
				& Sens (\%) & FPR/h & AUC & p-value & Sens (\%) & FPR/h & AUC & p-value & Sens (\%) & FPR/h & AUC & p-value & Sens (\%) & FPR/h & AUC & p-value & Sens (\%) & FPR/h & AUC & p-value \\
				\midrule
				Dog1 & 0.0 & \textbf{0.000} & 0.5000 & 1.000 & 50.0 & 0.275 & 0.6331 & 0.119 & 50.0 & 0.175 & 0.6197 & 0.083 & 50.0 & 0.163 & 0.6186 & 0.313 & \textbf{75.0} & 0.138 & \textbf{0.6653} & \textbf{0.015} \\
				Dog2 & 57.1 & \textbf{0.048} & 0.7065 & \textless0.001 & 100.0 & 0.216 & 0.8514 & \textless0.001 & 100.0 & 0.120 & 0.8528 & \textless0.001 & 100.0 & 0.132 & 0.8608 & \textless0.001 & \textbf{100.0} & 0.096 & \textbf{0.9089} & \textbf{\textless0.001} \\
				Dog3 & 8.3 & \textbf{0.000} & 0.5159 & 0.020 & 83.3 & 0.167 & 0.7713 & \textless0.001 & 91.7 & 0.117 & 0.7722 & \textless0.001 & 91.7 & 0.121 & 0.7884 & \textless0.001 & \textbf{91.7} & 0.054 & \textbf{0.8245} & \textbf{\textless0.001} \\
				Dog4 & 14.3 & \textbf{0.000} & 0.5630 & 0.097 & 78.6 & 0.269 & 0.6822 & \textless0.001 & 85.7 & 0.209 & 0.7127 & \textless0.001 & 85.7 & 0.201 & 0.7208 & \textless0.001 & \textbf{85.7} & 0.194 & \textbf{0.7471} & \textbf{\textless0.001} \\
				Dog5 & 40.0 & \textbf{0.040} & 0.6393 & 0.008 & 80.0 & 0.120 & 0.8176 & \textless0.001 & 80.0 & 0.053 & 0.8415 & \textless0.001 & 80.0 & 0.053 & 0.8383 & \textless0.001 & \textbf{100.0} & 0.080 & \textbf{0.9199} & \textbf{\textless0.001} \\
				\midrule
				Ave & 23.9 & \textbf{0.018} & 0.585 & - & 78.4 & 0.209 & 0.751 & - & 81.5 & 0.135 & 0.760 & - & 81.5 & 0.134 & 0.765 & - & \textbf{90.5} & 0.112 & \textbf{0.813} & \textbf{-} \\
				\bottomrule
			\end{tabular}
			\label{tab10}
		}
	\end{center}
\end{table*}

\begin{table*}[htbp]
	\caption{Performance of Multi-scale CNN on Kaggle database.}
	\begin{center}
		\resizebox{\textwidth}{!}{
			\begin{tabular}{ccccccccccccccccccccc}
				\toprule
				Patient & \multicolumn{4}{c}{Original} & \multicolumn{4}{c}{Down-sampling} & \multicolumn{4}{c}{Sliding windows} & \multicolumn{4}{c}{Recombination} & \multicolumn{4}{c}{\textbf{DiffEEG}} \\
				\cline{2-21}
				& Sens (\%) & FPR/h & AUC & p-value & Sens (\%) & FPR/h & AUC & p-value & Sens (\%) & FPR/h & AUC & p-value & Sens (\%) & FPR/h & AUC & p-value & Sens (\%) & FPR/h & AUC & p-value \\
				\midrule
				Dog1 & 0.0 & \textbf{0.000} & 0.5000 & 1.000 & 50.0 & 0.187 & 0.6305 & \textless0.001 & 50.0 & 0.225 & 0.6177 & \textless0.001 & 50.0 & 0.200 & 0.6481 & \textless0.001 & \textbf{75.0} & 0.138 & \textbf{0.6900} & \textbf{\textless0.001} \\
				Dog2 & 85.7 & \textbf{0.012} & 0.8003 & \textless0.001 & 100.0 & 0.204 & 0.8728 & \textless0.001 & 100.0 & 0.144 & 0.8865 & \textless0.001 & 100.0 & 0.132 & 0.8927 & \textless0.001 & \textbf{100.0} & 0.096 & \textbf{0.9065} & \textbf{\textless0.001} \\
				Dog3 & 58.3 & \textbf{0.013} & 0.7197 & \textless0.001 & 91.7 & 0.116 & 0.7880 & \textless0.001 & 91.7 & 0.138 & 0.7700 & \textless0.001 & 83.3 & 0.100 & 0.7849 & \textless0.001 & \textbf{100.0} & 0.100 & \textbf{0.8300} & \textbf{\textless0.001} \\
				Dog4 & 71.4 & \textbf{0.097} & 0.6444 & \textless0.001 & 85.7 & 0.268 & 0.7188 & \textless0.001 & 92.9 & 0.306 & 0.7265 & \textless0.001 & 92.9 & 0.239 & 0.7343 & \textless0.001 & \textbf{92.9} & 0.230 & \textbf{0.7686} & \textbf{\textless0.001} \\
				Dog5 & 80.0 & \textbf{0.027} & 0.7651 & \textless0.001 & 100.0 & 0.066 & 0.8886 & \textless0.001 & 100.0 & 0.053 & 0.8716 & \textless0.001 & 100.0 & 0.053 & 0.8877 & \textless0.001 & \textbf{100.0} & 0.040 & \textbf{0.9147} & \textbf{\textless0.001} \\
				\midrule
				Ave & 59.1 & \textbf{0.030} & 0.686 & - & 85.5 & 0.168 & 0.780 & - & 86.9 & 0.173 & 0.774 & - & 85.2 & 0.145 & 0.790 & - & \textbf{93.6} & 0.121 & \textbf{0.822} & \textbf{-} \\
				\bottomrule
			\end{tabular}
			\label{tab11}
		}
	\end{center}
\end{table*}

\begin{table*}[htbp]
	\caption{Performance of Transformer on Kaggle database.}
	\begin{center}
		\resizebox{\textwidth}{!}{
			\begin{tabular}{ccccccccccccccccccccc}
				\toprule
				Patient & \multicolumn{4}{c}{Original} & \multicolumn{4}{c}{Down-sampling} & \multicolumn{4}{c}{Sliding windows} & \multicolumn{4}{c}{Recombination} & \multicolumn{4}{c}{\textbf{DiffEEG}} \\
				\cline{2-21}
				& Sens (\%) & FPR/h & AUC & p-value & Sens (\%) & FPR/h & AUC & p-value & Sens (\%) & FPR/h & AUC & p-value & Sens (\%) & FPR/h & AUC & p-value & Sens (\%) & FPR/h & AUC & p-value \\
				\midrule
				Dog1 & 0.0 & \textbf{0.000} & 0.5000 & 1.000 & 50.0 & 0.225 & 0.6013 & 0.106 & 50.0 & 0.150 & 0.5941 & \textless0.001 & 50.0 & 0.125 & 0.5962 & 0.068 & \textbf{75.0} & 0.138 & \textbf{0.6387} & \textbf{\textless0.001} \\
				Dog2 & 71.4 & \textbf{0.036} & 0.6576 & \textless0.001 & 100.0 & 0.156 & 0.8697 & \textless0.001 & 100.0 & 0.144 & 0.8972 & \textless0.001 & 100.0 & 0.168 & 0.8508 & \textless0.001 & \textbf{100.0} & 0.120 & \textbf{0.9166} & \textbf{\textless0.001} \\
				Dog3 & 25.0 & \textbf{0.013} & 0.5398 & \textless0.001 & 83.3 & 0.100 & 0.7607 & \textless0.001 & 91.7 & 0.104 & 0.7435 & \textless0.001 & 91.7 & 0.104 & 0.7644 & \textless0.001 & \textbf{91.7} & 0.033 & \textbf{0.8155} & \textbf{\textless0.001} \\
				Dog4 & 0.0 & \textbf{0.000} & 0.5000 & 1.000 & 78.6 & 0.470 & 0.6449 & \textless0.001 & 78.6 & 0.300 & 0.6800 & \textless0.001 & 78.6 & 0.447 & 0.6391 & \textless0.001 & \textbf{85.7} & 0.201 & \textbf{0.7243} & \textbf{\textless0.001} \\
				Dog5 & 80.0 & \textbf{0.040} & 0.7522 & \textless0.001 & 100.0 & 0.133 & 0.8166 & \textless0.001 & 100.0 & 0.067 & 0.8478 & \textless0.001 & 100.0 & 0.080 & 0.8756 & \textless0.001 & \textbf{100.0} & 0.053 & \textbf{0.9179} & \textbf{\textless0.001} \\			
				\midrule
				Ave & 35.3 & \textbf{0.018} & 0.590 & - & 82.4 & 0.217 & 0.739 & - & 84.1 & 0.153 & 0.753 & - & 84.1 & 0.185 & 0.745 & - & \textbf{90.5} & 0.109 & \textbf{0.803} & \textbf{-} \\
				\bottomrule
			\end{tabular}
			\label{tab12}
		}
	\end{center}
\end{table*}

It is clear that all the classification models using the augmented data of DiffEEG achieve higher Sens and AUC than those of the models using just the original data. Among the approaches to solve the problem of imbalanced data, our proposed model achieves the highest Sens, AUC and the lowest FPR on all classifiers. Compared with the sliding windows and recombination methods, DiffEEG improves Sens by an average of 3.64\% and 4.06\%, AUC by an average of 0.043 and 0.041, and reduces FPR by an average of 0.027/h and 0.032/h on the CHB-MIT database. As for the Kaggle database, it improves Sens by an average of 7.70\% and 6.90\%, AUC by an average of 0.042 and 0.036, and reduces FPR by an average of 0.034/h and 0.037/h. Besides, our approach has increased the number of subjects whose improvement over chance is statistically significant, at a significant level of 0.05.

Among the five classification network, Multi-scale CNN wins the best performance. The combination of DiffEEG and Multi-scale CNN obtains an average Sens, FPR, AUC of 95.4\%, 0.051/h, 0.932 on the CHB-MIT database and an average Sens, FPR, AUC of 93.6\%, 0.121/h, 0.822 on the Kaggle database. 

To further measure our proposed model, we compare our results with several SOTA methods that used the CHB-MIT and Kaggle database. The statistical comparison is performed by counting the number of patients whose improvement over chance is statistically significant at a given confidence interval. The significance level $\alpha$ is set to 0.05 and 0.01, representing a confidence interval of 95\% and 99\% respectively. The comparison results are listed in Table \ref{tab13}. It is obvious that our method achieves the optimal performance on the CHB-MIT database, and comparable performance to the SOTA approaches on the Kaggle database. Besides, our work has acquired statistically significant improvement over chance for all the subjects in both databases. For the individuals without statistical significance in these SOTA methods, our work can achieve statistical significance on them, except for Pat7 who does not meet our selection criteria. 

\begin{table*}[htbp]
	\caption{Comparison to existing methods on CHB-MIT and Kaggle database.}
	\begin{center}
		\resizebox{\textwidth}{!}{
			\begin{tabular}{cccccccccccc}
				\toprule
				Authors & Database & Features & Classifier & \makecell{No. of \\ seizures} & \makecell{No. of \\ subjects} & Sens(\%) & FPR/h & AUC & \makecell{Interictal-Preictal \\ distance (min)} & \makecell{p-value over chance \\ at \textit{$\alpha$} = 0.05} & \makecell{p-value over chance \\ at \textit{$\alpha$} = 0.01} \\
				\midrule
				Troung et al. 2018 \cite{truong2018convolutional} & CHB-MIT & STFT & CNN & 64 & 13 & 81.2 & 0.16 & NR & 240-30 & 12/13 (miss Pat9) & 10/13 (miss Pat5, Pat9, Pat10) \\
				Ozcan et al. 2019 \cite{ozcan2019seizure} & CHB-MIT & Spectral power & 3D CNN & 77 & 16 & 85.7 & 0.096 & NR & 240-60 & 15/16 (miss Pat18) & 13/16 (miss Pat7, Pat10, Pat18) \\
                 & & Statistical moments & & & & & & & & \\
                 & & Hjorth parameters & & & & & & & & \\
				Zhang et al. 2020 \cite{zhang2019epilepsy} & CHB-MIT & Common spatial & CNN & 156 & 23 & 92 & 0.12 & 0.900 & 30-30 & NR & NR \\
                 & & pattern statistics & & & & & & & & \\
				Yang et al. 2021 \cite{yang2021effective} & CHB-MIT & STFT & RDANet & 64 & 13 & 89.3 & NR & 0.913 & 240-30 & NR & NR \\
				Li et al. 2022 \cite{li2022patient} & CHB-MIT & raw EEG signals & FB-CapsNet & 105 & 19 & 93.4 & 0.096 & 0.928 & 240-30 & 19/19 & 18/19 (miss Pat17) \\
				Zhao et al. 2022 \cite{zhao2022patient} & CHB-MIT & raw EEG signals & AddNet-SCL & 105 & 19 & 93.0 & 0.094 & 0.929 & 240-30 & 19/19 & 18/19 (miss Pat17) \\
				\textbf{Our work} & CHB-MIT & raw EEG signals & \makecell{CNN} & 69 & 13 & \textbf{95.4} & \textbf{0.051} & \textbf{0.932} & 240-30 & 13/13 & 13/13 \\
                 & & generated EEG & & & & & & & & \\
                \midrule
                Troung et al. 2018 \cite{truong2018convolutional} & Kaggle & STFT & CNN & 42 & 5 & 73.4 & 0.186 & NR & 240-60 & 4/5 (miss Dog1) & 4/5 (miss Dog1) \\
                Daoud et al. 2019 \cite{daoud2019efficient} & Kaggle & raw EEG signals & DCNN+Bi-LSTM & 42 & 5 & 81.5 & 0.167 & 0.805 & 240-60 & 4/5 (miss Dog1) & 4/5 (miss Dog1) \\
                Xu et al. 2020 \cite{xu2020end} & Kaggle & raw EEG signals & CNN & 42 & 5 & 80.9 & 0.134 & 0.808 & 240-60 & 4/5 (miss Dog1) & 4/5 (miss Dog1) \\
                Chen et al. 2021 \cite{chen2021seizure} & Kaggle & STFT & CNN & 42 & 5 & 80 & 0.372 & 0.828 & 240-60 & NR & NR \\
                Zhao et al. 2022 \cite{zhao2022patient} & Kaggle & raw EEG signals & SCL-AddNets & 42 & 5 & 89.1 & 0.12 & 0.831 & 240-60 & 5/5 & 5/5 \\
                Gao et al. 2023 \cite{gao2023self} & Kaggle & raw EEG signals & ProtoPNet & 42 & 5 & 88.6 & 0.146 & 0.764 & 240-60 & 5/5 & 5/5 \\
                \textbf{Our work} & Kaggle & raw EEG signals & \makecell{CNN} & 42 & 5 & \textbf{93.6} & \textbf{0.121} & \textbf{0.822} & 240-60 & 5/5 & 5/5 \\
                 & & generated EEG & & & & & & & & \\
				\bottomrule
			\end{tabular}
			\label{tab13}
		}
	\end{center}
\end{table*}

\begin{table}
	\caption{T-test between real and augmented data.}
	\begin{center}
		\resizebox{0.45\textwidth}{!}{
			\begin{tabular}{ccccccc}
				\toprule
				Subjects & Variance & \textit{$\delta$} band & \textit{$\theta$} band & \textit{$\alpha$} band & \textit{$\beta$} band & \textit{$\gamma$} band \\
				\midrule
				Pat1 & 0.970 & 0.811 & 0.419 & \textless0.001 & \textless0.001 & \textless0.001 \\
				Pat2 & 0.911 & 0.516 & 0.714 & 0.980 & 0.480 & \textless0.001 \\
				Pat3 & 0.987 & 0.898 & 0.027 & 0.893 & 0.813 & 0.722 \\
				Pat5 & 0.984 & \textless0.001 & \textless0.001 & 0.305 & \textless0.001 & 0.127 \\
				Pat9 & 0.072 & \textless0.001 & \textless0.001 & \textless0.001 & 0.068 & \textless0.001 \\
				Pat10 & 0.082 & 0.005 & 0.134 & \textless0.001 & \textless0.001 & \textless0.001 \\
                Pat13 & 0.503 & 0.063 & \textless0.001 & 0.059 & \textless0.001 & \textless0.001 \\
                Pat17 & 0.953 & 0.004 & 0.992 & 0.953 & 0.052 & \textless0.001 \\
                Pat18 & 0.870 & 0.347 & 0.094 & 0.949 & 0.035 & 0.514 \\
                Pat19 & 0.110 & 0.080 & 0.218 & 0.003 & 0.017 & 0.001 \\
                Pat20 & 0.108 & 0.310 & 0.188 & \textless0.001 & \textless0.001 & \textless0.001 \\
                Pat21 & 0.050 & 0.594 & 0.050 & \textless0.001 & \textless0.001 & \textless0.001 \\
                Pat23 & 0.582 & \textless0.001 & 0.360 & 0.840 & \textless0.001 & \textless0.001 \\
                \midrule
                Dog1 & 0.786 & \textless0.001 & \textless0.001 & 0.011 & 0.407 & \textless0.001 \\
                Dog2 & 0.190 & \textless0.001 & \textless0.001 & 0.164 & \textless0.001 & \textless0.001 \\
                Dog3 & 0.033 & \textless0.001 & \textless0.001 & 0.032 & \textless0.001 & \textless0.001 \\
                Dog4 & 0.101 & \textless0.001 & \textless0.001 & 0.109 & \textless0.001 & \textless0.001 \\
                Dog5 & 0.397 & \textless0.001 & 0.058 & \textless0.001 & \textless0.001 & \textless0.001 \\
				\bottomrule
			\end{tabular}
   }
			\label{tab14}
	\end{center}
\end{table}

\section{Discussion}\label{Disc}
Despite decades of development in medical technology, the collection of high-quality EEG signals still demands expensive medical devices and plenty of expert time \cite{rasheed2021generative}. It is impractical to increase the amount of preictal EEG data through continuous collection. Data augmentation is the process that generates new samples to augment a small or imbalanced database by converting existing samples. Exposing the model to various augmented data could raise the precision and stability of classifier, making it more robust to changes \cite{lashgari2020data}. According to the results in Table \ref{tab3} $\sim$ Table \ref{tab12}, it can be observed that all the classification models using the augmented data generated by DiffEEG achieve higher Sens and AUC than the models using just the original data, indicating that our method can improve the predictive precision of epilepsy. However, the models using the original data exhibited lower FPR. This is because the interictal samples in the original data are far more than the preictal data, leading to the overfitting of the model. As a result, the model can accurately identify the interictal samples but struggles to recognize the preictal samples. Besides, compared with the existing methods for addressing the imbalanced data problem, we can conclude that the DiffEEG has obvious advantage over them. There has been a notable enhancement in the performance of seizure prediction, proving that the data generated by DiffEEG has high quality and diversity. From a statistical perspective, the augmented data can increase the number of subjects whose improvement over chance is statistically significant, suggesting that our DA technique is capable of enhancing the reliability of the model’s predictions. In practice, the distribution of EEG signals varies over time due to the changes in external environment and mental states of patients. The epileptic representation in different preictal time varies as well if the location or mode of abnormal discharge is different \cite{qi2021learning}. When the data cluster of an upcoming seizure has few nearby distribution with the existing clusters, the seizure will be difficult to predict. Based on the distribution learned during the diffusion process, the data produced by DiffEEG can narrow the distance between different clusters and provide additional information to classifiers. Instead, the distribution of data generated by sliding windows or recombination is restricted by the original data and cannot be fully explored, so their improvement to prediction performance is limited. 

By comparing the results of five classification networks, it is clear that no matter what DA methods we used, the performance of Multi-scale CNN was the best. SVM is a powerful machine learning classifier, but its performance typically relies on the handcrafted feature engineering. The manually extracted features are hard to capture all the complex representations from raw data, making the performance of SVM inferior to that of deep learning networks. Although the blocks in Spatio-temporal MLP can extract information within and between channels, the fully connected structure of MLPs determines that the efficiency of data processing is not high. The lack of weight sharing mechanism also causes a huge amount of parameters, making it difficult to converge to optimal result. The structure of EEGNet is lightweight and computationally efficient. The multi-layer 1D convolution enables the model to effectively extract the time-frequency information. However, EEGNet lacks consideration of the inter-channel correlation, leading to the insufficient extraction of spatial features. For Transformer, the input embedding and position encoding can capture the global information of data. The attention module can learn the relation between features and strengthen the role of key information. For Multi-scale CNN, the local perception ability of convolution makes it easier to extract local information. By the use of dilated convolution, spatial and temporal features of different scales can be extracted, providing multi-scale information for seizure prediction. In the meantime, the model also utilizes attention mechanism as Transformer to assign weights to features and focus on the critical representation. Therefore, the Multi-scale CNN network obtains the best effects. SVM is a widely used traditional classification method. MLP, CNN and Transformer are the most representative frameworks in deep learning and almost all classifiers are based on them. As our proposed DiffEEG shows superiority on these networks, it is proved that DiffEEG can be utilized in almost any other classification network to substantially improve the seizure prediction performance.

\begin{figure}[htbp]
	\centerline{\includegraphics[width=0.45\textwidth]{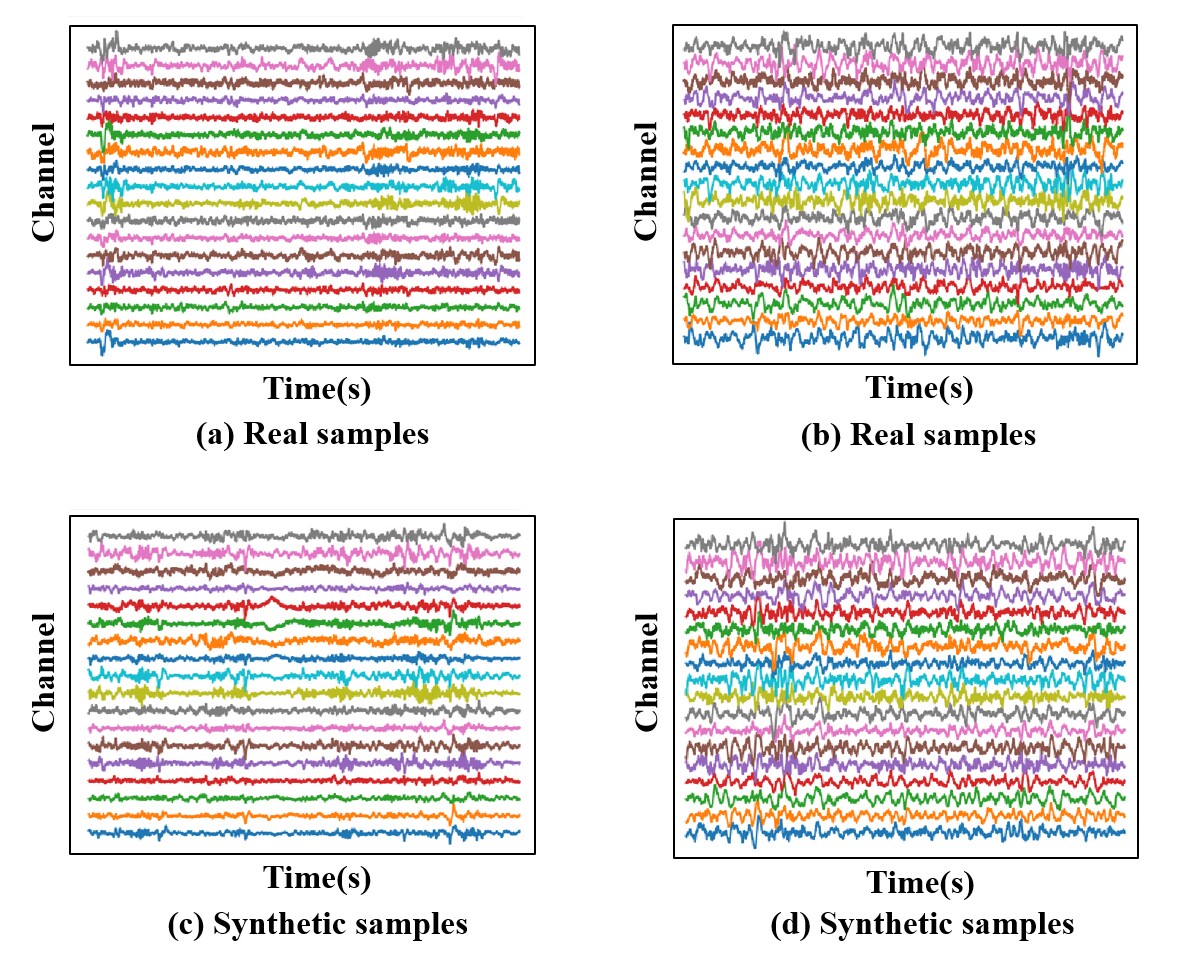}}
	\caption{Real and synthetic samples of 18 channels and 30-s duration. (a)(b) are real samples and (c)(d) are synthetic samples. The amplitudes of all samples are normalized to the same range.}
	\label{fig8}
\end{figure}

\begin{figure}[htbp]
	\centerline{\includegraphics[width=0.45\textwidth]{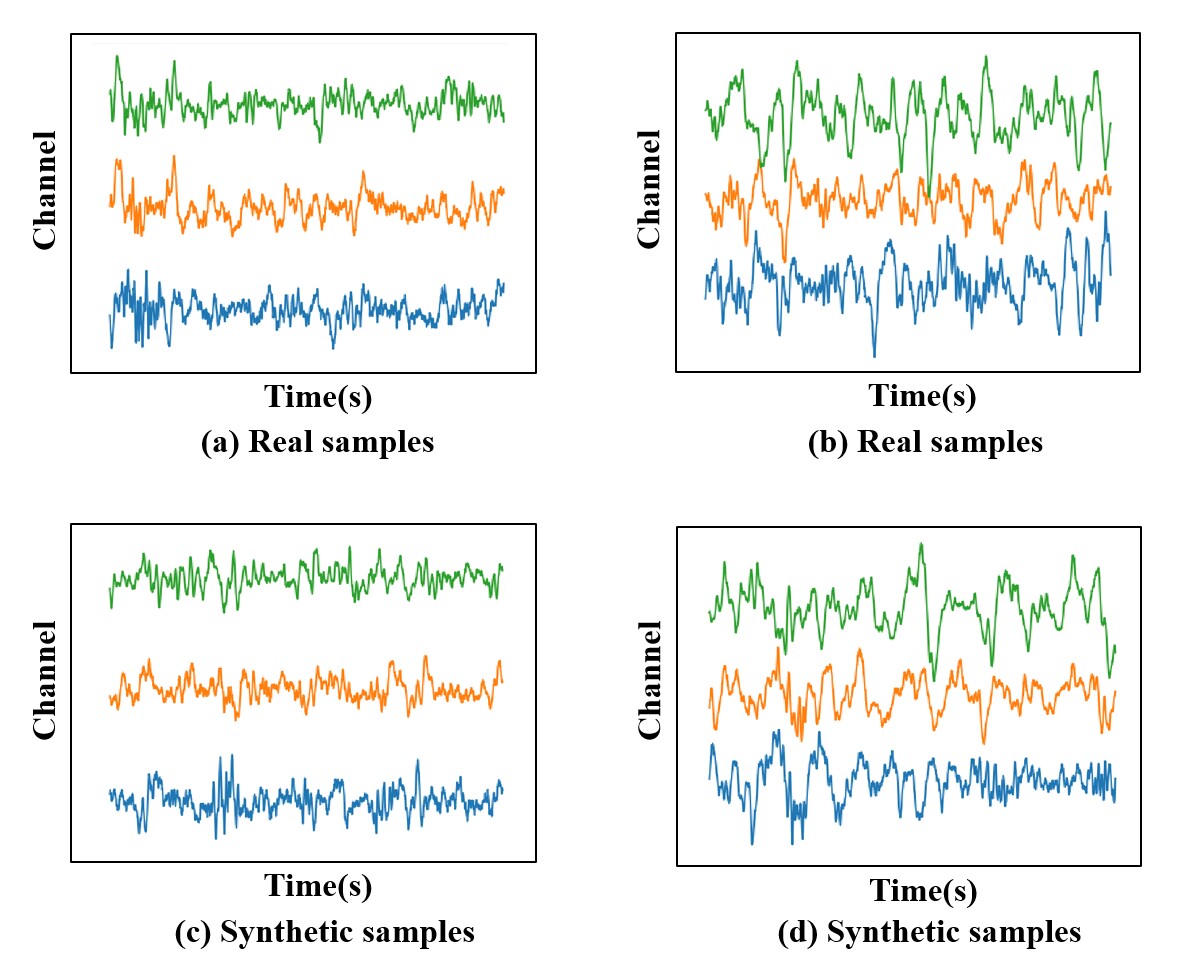}}
	\caption{The enlarged real and synthetic signals of 3 channels and 10-s duration. (a)(b) are from real samples and (c)(d) are from synthetic samples. The amplitudes of all samples are normalized to the same range.}
	\label{fig9} 
\end{figure}

The generation of multichannel EEG signals has been a challenge to researchers for a long time. Therefore, the existing generative models either generate the feature maps of EEG signals \cite{rasheed2021generative} or generate each channel respectively before combining them \cite{xu2022multichannel}. This is done in case the model cannot fully learn the representations of multichannel EEG signals. Our proposed model, DiffEEG, can overcome the challenge and directly generate signals of the whole channels. The real and synthetic samples of 18 channels and 30-s duration generated by DiffEEG are shown in Fig. \ref{fig8}. To further inspect the details of the samples, we only select the first three channels of 10-s duration as shown in Fig. \ref{fig9}. Through visual inspection, we can see that the augmented data is not the simple superposition of existing data, but diversified data generated based on the learned distribution, which could provide additional information to improve the performance.

Furthermore, statistical tests are performed for the synthetic data to validate that the augmented data increase diversity and new information while preserving the characteristics of the genuine EEG data. As the augmented data are generated conditioned on the STFT spectrum of real data, a paired t-test can be appropriately performed for the augmented data and the corresponding real data. T-tests are usually carried out on the feature space instead of the high-dimensional time-domain data \cite{islam2021crash}. Hence, we select variance to reflect the amplitude distribution, and extract the spectral power of $\delta$, $\theta$, $\alpha$, $\beta$, $\gamma$ bands to give information about the energy variations. We utilized these 6 features to conduct the paired t-test at a significance level of 0.05. The p-values on each subject are shown in Table \ref{tab14}. Based on the theory of t-test, if the p-value is less than 0.05, there is considered to be a significant difference between the two sample groups; otherwise, there is no significant difference. From the results, it can be observed that the variance of both groups is not significantly different across all subjects except for Dog3, indicating that the amplitude distribution of the augmented data is similar to that of the real data for most patients. For each spectral feature, some of the patients exhibit significant difference between the real and augmented groups, some of the patients not. This reflects the diversity introduced by our DiffEEG model in the generated samples. The synthetic data retain the temporal rhythms of the original data while bringing in diverse variations in the frequency domain.

The DiffEEG model is trained with the guidance of STFT condition. Compared to directly using the raw EEG signal, the conditional model can generate samples with higher quality. Numerous articles have demonstrated the advantages of conditional diffusion models over unconditional ones. Kong et al. mentioned that generating audios in the time domain without conditional information is a challenging task \cite{kong2020diffwave}. Their conditional model achieved higher Mean Opinion Score (MOS) values than the unconditional model, indicating that the generated samples of the conditional model have higher quality. Ho et al. also showed that the sample quality and fidelity can be greatly improved by the conditional guidance \cite{dhariwal2021diffusion}. Certainly, the conditional models also have their limitations. Ho et al. explained that the presence of condition could reduce the diversity of generation, resulting in a trade-off between quality and diversity. In our EEG generation task, we prioritize the quality because the complicated distribution of EEG data is difficult to learn. The loss of authenticity in the augmented data will lead to a decline in the seizure prediction performance. Besides, we have made use of the recombined spectrograms to increase the diversity as mentioned in Section \ref{Train}.

Finally, seizure prediction experiments are carried out to test the contributions of the synthetic data. The results verify that the data generated by DiffEEG can greatly improve the prediction performance and beat the existing DA methods in various situations.

The outstanding effects achieved by DiffEEG on public datasets have revealed multiple potential impacts on the clinical practice of seizure prediction.
\begin{itemize}
	\item Rapidly establishing high-performance and generalizable epilepsy prediction models for patients. In clinical practice, the insufficient preictal data often restricts the performance of seizure prediction models \cite{xu2022multichannel}. Relying on collecting more preictal data to build high-performance models is extremely time-consuming. With DiffEEG, preictal data with additional information can be generated in a short time, enhancing the predictive effect of the model. Moreover, the strong generalization capability of the synthetic data allows them to be widely applicable to various prediction models.
	\item Assisting doctors in studying biomarkers for epilepsy prediction and enhancing the interpretability of predictions. It is challenging to capture the EEG biomarkers before seizures because of the complex spatial-temporal dynamics present in the epileptic brain \cite{li2021spatio}. The DiffEEG model can fully explore the feature distribution of EEG signals and generates data with distinct epileptic representation. These data provide a possible reference for doctors in identifying biomarkers for seizure prediction, improving the interpretability of deep learning-based prediction models.  
	\item Protecting the data privacy of patients. In the medical field, EEG data is typically sensitive information. Utilizing synthetic samples for data augmentation can reduce the reliance on original data, thereby lowering the risks of data leakage and privacy concerns. 
\end{itemize}

Our work also presents vast application prospects in other EEG-based tasks. Issues such as data imbalance and scarcity are prevalent in EEG-based tasks. For instance, in sleep staging tasks, Stage N2 generally occupies a majority of the sleep time (45\%-55\%), while stage N1 only accounts for very little time (2\%-5\%) \cite{fan2020eeg}. In emotion recognition tasks, the AMIGOS dataset exhibits a higher arousal ratings, whereas in the DEAP dataset, the proportion of positive emotions is larger than that of negative ones \cite{zhang2024beyond}. In the diagnosis of brain diseases such as depression, Alzheimer's disease and mild cognitive impairment, it is typical to face a lack of data due to the equipment requirements, time constraints and privacy concerns. The EEG signals for all these tasks exhibit significantly different feature distribution. However, DiffEEG can leverage the STFT to provide time-frequency information, guiding the model to learn task-specific representations for each distribution. Therefore, our approach enables the generation of diverse and high-quality EEG samples for data augmentation under various tasks, raising the generalization, accuracy, and robustness of these tasks.

Despite the promising influence and applications demonstrated by DiffEEG, it is important to acknowledge its potential limitations and pursue the directions for future research. One of the primary limitations is the slow training process. Although we speed up the training process by randomly selecting the diffusion steps, it still takes plenty of time to train the model. In future work, the input block could be modified to transform the multichannel EEG data into latent space. By performing the diffusion process in a low-dimensional latent space, the computation cost and training time can be greatly reduced. Another limitation is the lack of cross-subject capabilities. To ensure the generation quality, DiffEEG is trained under a subject-specific strategy. When a new patient arrives, we need to retrain a model rather than utilizing the existing data and models, which is both time-consuming and resource-wasting. In the future, the structure of DiffEEG could be improved for pretraining on a large dataset. In this way, we only need to fine-tune the pretrained model to fit the data distribution of each patient, enabling the model with few-shot learning capability.

In addition to the inherent limitations of the DiffEEG itself, there are several challenges and considerations when integrating it into real-world applications. The first obstacle comes from the acquisition quality of the EEG signals. In real-world scenarios, factors such as the loose or offset of the collecting electrodes, the patient's chewing and vigorous movement can introduce severe noise in the EEG signals \cite{karpov2021noise}. The noise interference may mask or distort the epileptic representation, hindering the model from learning the distribution characteristics of epileptic EEG data. We cannot guarantee that a model trained on low-quality EEG signals is capable of producing high-quality EEG data. The second challenge is the difficulty in continuously optimizing the model. Patients employing the epilepsy prediction system still experience seizures, and the consequent preictal data could be utilized to update and enhance the DiffEEG model. However, retraining the model with all the data each time new data arrives is too costly, and training the model solely on the new data could lead to the catastrophic forgetting problem. Hence, integrating incremental learning into DiffEEG is worth considering.  At last, the highly realistic content generated by DiffEEG can raise ethical concerns. The model has the potential to create deceptive EEG signals including deepfakes and misinformation \cite{katirai2024situating}. Therefore, ensuring responsible use and preventing malicious applications are ongoing challenges.

\section{Conclusion}\label{Conc}
In this article, we propose a novel and effective data augmentation method by the use of diffusion model. We conduct contrast experiments with just using the original data and with three existing methods of solving the imbalanced data problem: down-sampling, sliding windows and recombination. We evaluate the prediction performance on five representative classifiers to verify the effectiveness and generality of DiffEEG. The results prove that the data generated by DiffEEG achieve better performance on all five classification networks. Among them, the Multi-scale CNN utilizing the data augmented by DiffEEG obtained the best performance, with an average Sens, FPR, AUC of 95.4\%, 0.051/h, 0.932 on the CHB-MIT database and an average Sens, FPR, AUC of 93.6\%, 0.121/h, 0.822 on the Kaggle database. Both results reached the SOTA level. To the best of our knowledge, our work is the first diffusion model used in epileptic EEG generation, creating a new and effective way to solve the imbalanced data. This model can be widely applied to all kinds of seizure prediction classifiers to improve the classification performance significantly.

\bibliographystyle{IEEEtran}
\bibliography{ustc.bib}

\begin{thebibliography}{10}
\providecommand{\url}[1]{#1}
\csname url@samestyle\endcsname
\providecommand{\newblock}{\relax}
\providecommand{\bibinfo}[2]{#2}
\providecommand{\BIBentrySTDinterwordspacing}{\spaceskip=0pt\relax}
\providecommand{\BIBentryALTinterwordstretchfactor}{4}
\providecommand{\BIBentryALTinterwordspacing}{\spaceskip=\fontdimen2\font plus
\BIBentryALTinterwordstretchfactor\fontdimen3\font minus \fontdimen4\font\relax}
\providecommand{\BIBforeignlanguage}[2]{{%
\expandafter\ifx\csname l@#1\endcsname\relax
\typeout{** WARNING: IEEEtran.bst: No hyphenation pattern has been}%
\typeout{** loaded for the language `#1'. Using the pattern for}%
\typeout{** the default language instead.}%
\else
\language=\csname l@#1\endcsname
\fi
#2}}
\providecommand{\BIBdecl}{\relax}
\BIBdecl

\bibitem{fisher2014ilae}
R.~S. Fisher, C.~Acevedo, A.~Arzimanoglou, A.~Bogacz, J.~H. Cross, C.~E. Elger, J.~Engel~Jr, L.~Forsgren, J.~A. French, M.~Glynn \emph{et~al.}, ``Ilae official report: a practical clinical definition of epilepsy,'' \emph{Epilepsia}, vol.~55, no.~4, pp. 475--482, 2014.

\bibitem{world2019epilepsy}
W.~H. Organization \emph{et~al.}, \emph{Epilepsy: a public health imperative}.\hskip 1em plus 0.5em minus 0.4em\relax World Health Organization, 2019.

\bibitem{zeng2020hierarchy}
D.~Zeng, K.~Huang, C.~Xu, H.~Shen, and Z.~Chen, ``Hierarchy graph convolution network and tree classification for epileptic detection on electroencephalography signals,'' \emph{IEEE Transactions on Cognitive and Developmental Systems}, vol.~13, no.~4, pp. 955--968, 2020.

\bibitem{maimaiti2022overview}
B.~Maimaiti, H.~Meng, Y.~Lv, J.~Qiu, Z.~Zhu, Y.~Xie, Y.~Li, W.~Zhao, J.~Liu, M.~Li \emph{et~al.}, ``An overview of eeg-based machine learning methods in seizure prediction and opportunities for neurologists in this field,'' \emph{Neuroscience}, vol. 481, pp. 197--218, 2022.

\bibitem{xiao2021automatic}
L.~Xiao, C.~Li, Y.~Wang, J.~Chen, W.~Si, C.~Yao, X.~Li, C.~Duan, and P.-A. Heng, ``Automatic localization of seizure onset zone from high-frequency seeg signals: A preliminary study,'' \emph{IEEE Journal of Translational Engineering in Health and Medicine}, vol.~9, pp. 1--10, 2021.

\bibitem{cao2019epileptic}
J.~Cao, J.~Zhu, W.~Hu, and A.~Kummert, ``Epileptic signal classification with deep eeg features by stacked cnns,'' \emph{IEEE Transactions on Cognitive and Developmental Systems}, vol.~12, no.~4, pp. 709--722, 2019.

\bibitem{kuhlmann2018seizure}
L.~Kuhlmann, K.~Lehnertz, M.~P. Richardson, B.~Schelter, and H.~P. Zaveri, ``Seizure prediction—ready for a new era,'' \emph{Nature Reviews Neurology}, vol.~14, no.~10, pp. 618--630, 2018.

\bibitem{daoud2019efficient}
H.~Daoud and M.~A. Bayoumi, ``Efficient epileptic seizure prediction based on deep learning,'' \emph{IEEE Transactions on Biomedical Circuits and Systems}, vol.~13, no.~5, pp. 804--813, 2019.

\bibitem{chen2022toward}
X.~Chen, C.~Li, A.~Liu, M.~J. McKeown, R.~Qian, and Z.~J. Wang, ``Toward open-world electroencephalogram decoding via deep learning: A comprehensive survey,'' \emph{IEEE Signal Processing Magazine}, vol.~39, no.~2, pp. 117--134, 2022.

\bibitem{chisci2010real}
L.~Chisci, A.~Mavino, G.~Perferi, M.~Sciandrone, C.~Anile, G.~Colicchio, and F.~Fuggetta, ``Real-time epileptic seizure prediction using ar models and support vector machines,'' \emph{IEEE Transactions on Biomedical Engineering}, vol.~57, no.~5, pp. 1124--1132, 2010.

\bibitem{ozcan2019seizure}
A.~R. Ozcan and S.~Erturk, ``Seizure prediction in scalp eeg using 3d convolutional neural networks with an image-based approach,'' \emph{IEEE Transactions on Neural Systems and Rehabilitation Engineering}, vol.~27, no.~11, pp. 2284--2293, 2019.

\bibitem{cao2021epileptic}
J.~Cao, D.~Hu, Y.~Wang, J.~Wang, and B.~Lei, ``Epileptic classification with deep-transfer-learning-based feature fusion algorithm,'' \emph{IEEE Transactions on Cognitive and Developmental Systems}, vol.~14, no.~2, pp. 684--695, 2021.

\bibitem{li2023spatio}
C.~Li, C.~Shao, R.~Song, G.~Xu, X.~Liu, R.~Qian, and X.~Chen, ``Spatio-temporal mlp network for seizure prediction using eeg signals,'' \emph{Measurement}, vol. 206, p. 112278, 2023.

\bibitem{gao2022pediatric}
Y.~Gao, X.~Chen, A.~Liu, D.~Liang, L.~Wu, R.~Qian, H.~Xie, and Y.~Zhang, ``Pediatric seizure prediction in scalp eeg using a multi-scale neural network with dilated convolutions,'' \emph{IEEE Journal of Translational Engineering in Health and Medicine}, vol.~10, pp. 1--9, 2022.

\bibitem{gao2022general}
Y.~Gao, A.~Liu, X.~Cui, R.~Qian, and X.~Chen, ``A general sample-weighted framework for epileptic seizure prediction,'' \emph{Computers in Biology and Medicine}, vol. 150, p. 106169, 2022.

\bibitem{cook2013prediction}
M.~J. Cook, T.~J. O'Brien, S.~F. Berkovic, M.~Murphy, A.~Morokoff, G.~Fabinyi, W.~D'Souza, R.~Yerra, J.~Archer, L.~Litewka \emph{et~al.}, ``Prediction of seizure likelihood with a long-term, implanted seizure advisory system in patients with drug-resistant epilepsy: a first-in-man study,'' \emph{The Lancet Neurology}, vol.~12, no.~6, pp. 563--571, 2013.

\bibitem{ihle2012epilepsiae}
M.~Ihle, H.~Feldwisch-Drentrup, C.~A. Teixeira, A.~Witon, B.~Schelter, J.~Timmer, and A.~Schulze-Bonhage, ``Epilepsiae--a european epilepsy database,'' \emph{Computer Methods and Programs in Biomedicine}, vol. 106, no.~3, pp. 127--138, 2012.

\bibitem{branco2016survey}
P.~Branco, L.~Torgo, and R.~P. Ribeiro, ``A survey of predictive modeling on imbalanced domains,'' \emph{ACM Computing Surveys (CSUR)}, vol.~49, no.~2, pp. 1--50, 2016.

\bibitem{khan2017focal}
H.~Khan, L.~Marcuse, M.~Fields, K.~Swann, and B.~Yener, ``Focal onset seizure prediction using convolutional networks,'' \emph{IEEE Transactions on Biomedical Engineering}, vol.~65, no.~9, pp. 2109--2118, 2017.

\bibitem{truong2018convolutional}
N.~D. Truong, A.~D. Nguyen, L.~Kuhlmann, M.~R. Bonyadi, J.~Yang, S.~Ippolito, and O.~Kavehei, ``Convolutional neural networks for seizure prediction using intracranial and scalp electroencephalogram,'' \emph{Neural Networks}, vol. 105, pp. 104--111, 2018.

\bibitem{zhang2019epilepsy}
Y.~Zhang, Y.~Guo, P.~Yang, W.~Chen, and B.~Lo, ``Epilepsy seizure prediction on eeg using common spatial pattern and convolutional neural network,'' \emph{IEEE Journal of Biomedical and Health Informatics}, vol.~24, no.~2, pp. 465--474, 2019.

\bibitem{antoniou2017data}
A.~Antoniou, A.~Storkey, and H.~Edwards, ``Data augmentation generative adversarial networks,'' \emph{arXiv preprint arXiv:1711.04340}, 2017.

\bibitem{qi2021learning}
Y.~Qi, L.~Ding, Y.~Wang, and G.~Pan, ``Learning robust features from nonstationary brain signals by multiscale domain adaptation networks for seizure prediction,'' \emph{IEEE Transactions on Cognitive and Developmental Systems}, vol.~14, no.~3, pp. 1208--1216, 2021.

\bibitem{trabucco2023effective}
B.~Trabucco, K.~Doherty, M.~Gurinas, and R.~Salakhutdinov, ``Effective data augmentation with diffusion models,'' \emph{arXiv preprint arXiv:2302.07944}, 2023.

\bibitem{rasheed2021generative}
K.~Rasheed, J.~Qadir, T.~J. O’Brien, L.~Kuhlmann, and A.~Razi, ``A generative model to synthesize eeg data for epileptic seizure prediction,'' \emph{IEEE Transactions on Neural Systems and Rehabilitation Engineering}, vol.~29, pp. 2322--2332, 2021.

\bibitem{salimans2016improved}
T.~Salimans, I.~Goodfellow, W.~Zaremba, V.~Cheung, A.~Radford, and X.~Chen, ``Improved techniques for training gans,'' \emph{Advances in Neural Information Processing Systems}, vol.~29, 2016.

\bibitem{giannone2022few}
G.~Giannone, D.~Nielsen, and O.~Winther, ``Few-shot diffusion models,'' \emph{arXiv preprint arXiv:2205.15463}, 2022.

\bibitem{kong2020diffwave}
Z.~Kong, W.~Ping, J.~Huang, K.~Zhao, and B.~Catanzaro, ``Diffwave: A versatile diffusion model for audio synthesis,'' \emph{arXiv preprint arXiv:2009.09761}, 2020.

\bibitem{ho2020denoising}
J.~Ho, A.~Jain, and P.~Abbeel, ``Denoising diffusion probabilistic models,'' \emph{Advances in Neural Information Processing Systems}, vol.~33, pp. 6840--6851, 2020.

\bibitem{kebaili2023deep}
A.~Kebaili, J.~Lapuyade-Lahorgue, and S.~Ruan, ``Deep learning approaches for data augmentation in medical imaging: A review,'' \emph{Journal of Imaging}, vol.~9, no.~4, p.~81, 2023.

\bibitem{pinaya2022brain}
W.~H. Pinaya, P.-D. Tudosiu, J.~Dafflon, P.~F. Da~Costa, V.~Fernandez, P.~Nachev, S.~Ourselin, and M.~J. Cardoso, ``Brain imaging generation with latent diffusion models,'' in \emph{Deep Generative Models: Second MICCAI Workshop, DGM4MICCAI 2022, Held in Conjunction with MICCAI 2022, Singapore, September 22, 2022, Proceedings}.\hskip 1em plus 0.5em minus 0.4em\relax Springer, 2022, pp. 117--126.

\bibitem{chung2022mr}
H.~Chung, E.~S. Lee, and J.~C. Ye, ``Mr image denoising and super-resolution using regularized reverse diffusion,'' \emph{IEEE Transactions on Medical Imaging}, 2022.

\bibitem{chambon2022roentgen}
P.~Chambon, C.~Bluethgen, J.-B. Delbrouck, R.~Van~der Sluijs, M.~Po{\l}acin, J.~M.~Z. Chaves, T.~M. Abraham, S.~Purohit, C.~P. Langlotz, and A.~Chaudhari, ``Roentgen: Vision-language foundation model for chest x-ray generation,'' \emph{arXiv preprint arXiv:2211.12737}, 2022.

\bibitem{dhariwal2021diffusion}
P.~Dhariwal and A.~Nichol, ``Diffusion models beat gans on image synthesis,'' \emph{Advances in Neural Information Processing Systems}, vol.~34, pp. 8780--8794, 2021.

\bibitem{cao2022survey}
H.~Cao, C.~Tan, Z.~Gao, G.~Chen, P.-A. Heng, and S.~Z. Li, ``A survey on generative diffusion model,'' \emph{arXiv preprint arXiv:2209.02646}, 2022.

\bibitem{song2020denoising}
J.~Song, C.~Meng, and S.~Ermon, ``Denoising diffusion implicit models,'' \emph{arXiv preprint arXiv:2010.02502}, 2020.

\bibitem{khan2018new}
N.~A. Khan and S.~Ali, ``A new feature for the classification of non-stationary signals based on the direction of signal energy in the time--frequency domain,'' \emph{Computers in Biology and Medicine}, vol. 100, pp. 10--16, 2018.

\bibitem{yang2021effective}
X.~Yang, J.~Zhao, Q.~Sun, J.~Lu, and X.~Ma, ``An effective dual self-attention residual network for seizure prediction,'' \emph{IEEE Transactions on Neural Systems and Rehabilitation Engineering}, vol.~29, pp. 1604--1613, 2021.

\bibitem{simonyan2014very}
K.~Simonyan and A.~Zisserman, ``Very deep convolutional networks for large-scale image recognition,'' \emph{arXiv preprint arXiv:1409.1556}, 2014.

\bibitem{vaswani2017attention}
A.~Vaswani, N.~Shazeer, N.~Parmar, J.~Uszkoreit, L.~Jones, A.~N. Gomez, {\L}.~Kaiser, and I.~Polosukhin, ``Attention is all you need,'' \emph{Advances in Neural Information Processing Systems}, vol.~30, 2017.

\bibitem{oord2016wavenet}
A.~v.~d. Oord, S.~Dieleman, H.~Zen, K.~Simonyan, O.~Vinyals, A.~Graves, N.~Kalchbrenner, A.~Senior, and K.~Kavukcuoglu, ``Wavenet: A generative model for raw audio,'' \emph{arXiv preprint arXiv:1609.03499}, 2016.

\bibitem{dauphin2017language}
Y.~N. Dauphin, A.~Fan, M.~Auli, and D.~Grangier, ``Language modeling with gated convolutional networks,'' in \emph{International Conference on Machine Learning}.\hskip 1em plus 0.5em minus 0.4em\relax PMLR, 2017, pp. 933--941.

\bibitem{he2016deep}
K.~He, X.~Zhang, S.~Ren, and J.~Sun, ``Deep residual learning for image recognition,'' in \emph{Proceedings of the IEEE Conference on Computer Vision and Pattern Recognition}, 2016, pp. 770--778.

\bibitem{lawhern2018eegnet}
V.~J. Lawhern, A.~J. Solon, N.~R. Waytowich, S.~M. Gordon, C.~P. Hung, and B.~J. Lance, ``Eegnet: a compact convolutional neural network for eeg-based brain--computer interfaces,'' \emph{Journal of Neural Engineering}, vol.~15, no.~5, p. 056013, 2018.

\bibitem{hu2015removal}
J.~Hu, C.-s. Wang, M.~Wu, Y.-x. Du, Y.~He, and J.~She, ``Removal of eog and emg artifacts from eeg using combination of functional link neural network and adaptive neural fuzzy inference system,'' \emph{Neurocomputing}, vol. 151, pp. 278--287, 2015.

\bibitem{lu2020staging}
D.~Lu, S.~Bauer, V.~Neubert, L.~S. Costard, F.~Rosenow, and J.~Triesch, ``Staging epileptogenesis with deep neural networks,'' in \emph{Proceedings of the 11th ACM International Conference on Bioinformatics, Computational Biology and Health Informatics}, 2020, pp. 1--10.

\bibitem{liu2021swin}
Z.~Liu, Y.~Lin, Y.~Cao, H.~Hu, Y.~Wei, Z.~Zhang, S.~Lin, and B.~Guo, ``Swin transformer: Hierarchical vision transformer using shifted windows,'' in \emph{Proceedings of the IEEE/CVF International Conference on Computer Vision}, 2021, pp. 10\,012--10\,022.

\bibitem{barandela2004imbalanced}
R.~Barandela, R.~M. Valdovinos, J.~S. S{\'a}nchez, and F.~J. Ferri, ``The imbalanced training sample problem: Under or over sampling?'' in \emph{Structural, Syntactic, and Statistical Pattern Recognition: Joint IAPR International Workshops, SSPR 2004 and SPR 2004, Lisbon, Portugal, August 18-20, 2004. Proceedings}.\hskip 1em plus 0.5em minus 0.4em\relax Springer, 2004, pp. 806--814.

\bibitem{shoeb2009application}
A.~H. Shoeb, ``Application of machine learning to epileptic seizure onset detection and treatment,'' Ph.D. dissertation, Massachusetts Institute of Technology, 2009.

\bibitem{brinkmann2016crowdsourcing}
B.~H. Brinkmann, J.~Wagenaar, D.~Abbot, P.~Adkins, S.~C. Bosshard, M.~Chen, Q.~M. Tieng, J.~He, F.~Mu{\~n}oz-Almaraz, P.~Botella-Rocamora \emph{et~al.}, ``Crowdsourcing reproducible seizure forecasting in human and canine epilepsy,'' \emph{Brain}, vol. 139, no.~6, pp. 1713--1722, 2016.

\bibitem{maiwald2004comparison}
T.~Maiwald, M.~Winterhalder, R.~Aschenbrenner-Scheibe, H.~U. Voss, A.~Schulze-Bonhage, and J.~Timmer, ``Comparison of three nonlinear seizure prediction methods by means of the seizure prediction characteristic,'' \emph{Physica D: Nonlinear Phenomena}, vol. 194, no. 3-4, pp. 357--368, 2004.

\bibitem{zhang2023distilling}
Z.~Zhang, A.~Liu, Y.~Gao, X.~Cui, R.~Qian, and X.~Chen, ``Distilling invariant representations with domain adversarial learning for cross-subject children seizure prediction,'' \emph{IEEE Transactions on Cognitive and Developmental Systems}, 2023.

\bibitem{li2022patient}
C.~Li, Y.~Zhao, R.~Song, X.~Liu, R.~Qian, and X.~Chen, ``Patient-specific seizure prediction from electroencephalogram signal via multi-channel feedback capsule network,'' \emph{IEEE Transactions on Cognitive and Developmental Systems}, 2022.

\bibitem{zhao2022patient}
Y.~Zhao, C.~Li, X.~Liu, R.~Qian, R.~Song, and X.~Chen, ``Patient-specific seizure prediction via adder network and supervised contrastive learning,'' \emph{IEEE Transactions on Neural Systems and Rehabilitation Engineering}, vol.~30, pp. 1536--1547, 2022.

\bibitem{xu2020end}
Y.~Xu, J.~Yang, S.~Zhao, H.~Wu, and M.~Sawan, ``An end-to-end deep learning approach for epileptic seizure prediction,'' in \emph{2020 2nd IEEE International Conference on Artificial Intelligence Circuits and Systems (AICAS)}.\hskip 1em plus 0.5em minus 0.4em\relax IEEE, 2020, pp. 266--270.

\bibitem{chen2021seizure}
R.~Chen and K.~K. Parhi, ``Seizure prediction using convolutional neural networks and sequence transformer networks,'' in \emph{2021 43rd Annual International Conference of the IEEE Engineering in Medicine \& Biology Society (EMBC)}.\hskip 1em plus 0.5em minus 0.4em\relax IEEE, 2021, pp. 6483--6486.

\bibitem{gao2023self}
Y.~Gao, A.~Liu, L.~Wang, R.~Qian, and X.~Chen, ``A self-interpretable deep learning model for seizure prediction using a multi-scale prototypical part network,'' \emph{IEEE Transactions on Neural Systems and Rehabilitation Engineering}, vol.~31, pp. 1847--1856, 2023.

\bibitem{lashgari2020data}
E.~Lashgari, D.~Liang, and U.~Maoz, ``Data augmentation for deep-learning-based electroencephalography,'' \emph{Journal of Neuroscience Methods}, vol. 346, p. 108885, 2020.

\bibitem{xu2022multichannel}
Y.~Xu, J.~Yang, and M.~Sawan, ``Multichannel synthetic preictal eeg signals to enhance the prediction of epileptic seizures,'' \emph{IEEE Transactions on Biomedical Engineering}, vol.~69, no.~11, pp. 3516--3525, 2022.

\bibitem{islam2021crash}
Z.~Islam, M.~Abdel-Aty, Q.~Cai, and J.~Yuan, ``Crash data augmentation using variational autoencoder,'' \emph{Accident Analysis \& Prevention}, vol. 151, p. 105950, 2021.

\bibitem{li2021spatio}
Y.~Li, Y.~Liu, Y.-Z. Guo, X.-F. Liao, B.~Hu, and T.~Yu, ``Spatio-temporal-spectral hierarchical graph convolutional network with semisupervised active learning for patient-specific seizure prediction,'' \emph{IEEE Transactions on Cybernetics}, vol.~52, no.~11, pp. 12\,189--12\,204, 2021.

\bibitem{fan2020eeg}
J.~Fan, C.~Sun, C.~Chen, X.~Jiang, X.~Liu, X.~Zhao, L.~Meng, C.~Dai, and W.~Chen, ``Eeg data augmentation: towards class imbalance problem in sleep staging tasks,'' \emph{Journal of Neural Engineering}, vol.~17, no.~5, p. 056017, 2020.

\bibitem{zhang2024beyond}
Z.~Zhang, S.~Zhong, and Y.~Liu, ``Beyond mimicking under-represented emotions: Deep data augmentation with emotional subspace constraints for eeg-based emotion recognition,'' in \emph{Proceedings of the AAAI Conference on Artificial Intelligence}, vol.~38, no.~9, 2024, pp. 10\,252--10\,260.

\bibitem{karpov2021noise}
O.~E. Karpov, V.~V. Grubov, V.~A. Maksimenko, N.~Utaschev, V.~E. Semerikov, D.~A. Andrikov, and A.~E. Hramov, ``Noise amplification precedes extreme epileptic events on human eeg,'' \emph{Physical Review E}, vol. 103, no.~2, p. 022310, 2021.

\bibitem{katirai2024situating}
A.~Katirai, N.~Garcia, K.~Ide, Y.~Nakashima, and A.~Kishimoto, ``Situating the social issues of image generation models in the model life cycle: a sociotechnical approach,'' \emph{AI and Ethics}, pp. 1--18, 2024.

\end{thebibliography}

\vfill
\end{CJK}
\end{document}